\documentclass[journal]{IEEEtran}

\usepackage{amsmath,amssymb,amsthm}
\usepackage{booktabs}
\usepackage{array}
\usepackage{multirow}
\usepackage{xspace}
\usepackage{graphicx}
\usepackage{url}
\usepackage[hidelinks]{hyperref}
\usepackage[capitalize,nameinlink]{cleveref}
\hypersetup{pdftitle={Fresh Keys, Stale Secrets: Numeric Leakage across Linked LLM Prompts},pdfauthor={Jian Li, Qinglin Zhao, Shuoyao Wang, Jincheng Cai, Zhifeng Hao, Naiqi Wu}}
\crefname{appendixsec}{Section}{Sections}
\Crefname{appendixsec}{Section}{Sections}
\newcommand{\R}{\mathbb{R}}
\newcommand{\E}{\mathbb{E}}
\newcommand{\Prob}{\mathbb{P}}
\newcommand{\ind}{\mathbf{1}}
\newcommand{\indep}{\perp\!\!\!\perp}
\newcommand{\Fset}{\mathcal{F}}
\newcommand{\Xset}{\mathcal{X}}

\newcommand{\Done}{D1\xspace}
\newcommand{\Donep}{D1$'$\xspace}
\newcommand{\Dtwo}{D2\xspace}
\newcommand{\Dthree}{D3\xspace}
\newcommand{\Dthreeo}{D3o\xspace}
\newcommand{\preempt}{Pr$\varepsilon\varepsilon$mpt\xspace}
\newcommand{\Aplus}{A1+\xspace}
\newcommand{\Azero}{A0\xspace}
\newcommand{\pp}{\,pp\xspace}

\newcommand{\Description}[1]{}
\newtheorem{theorem}{Theorem}
\newtheorem{proposition}[theorem]{Proposition}
\newtheorem{lemma}[theorem]{Lemma}
\newtheoremstyle{restated}{\topsep}{\topsep}{\itshape}{}{\bfseries}{.}{ }{\thmnote{#3}}
\theoremstyle{restated}

\theoremstyle{plain}

\begin{document}

\title{Fresh Keys, Stale Secrets: Numeric Leakage across Linked LLM Prompts}

\author{Jian Li,~\IEEEmembership{Member,~IEEE,}
        Qinglin~Zhao,~\IEEEmembership{Senior Member,~IEEE,}
        Shuoyao~Wang,~\IEEEmembership{Senior Member,~IEEE,}
        Jincheng~Cai,
        Zhifeng~Hao,~\IEEEmembership{Senior Member,~IEEE,}
        Naiqi~Wu,~\IEEEmembership{Fellow,~IEEE} 
\IEEEcompsocitemizethanks{
\IEEEcompsocthanksitem This work has been submitted to the IEEE for possible publication. Copyright may be transferred without notice, after which this version may no longer be accessible.
\IEEEcompsocthanksitem This work was supported in part by the National Natural Science Foundation of China under Project 62506082 and in part by the Guangdong Basic and Applied Basic Research Foundation under Project 2025A1515010164. \textit{(Corresponding author: Jian Li.)}
\IEEEcompsocthanksitem Jian Li is with the School of Advanced Manufacturing, Guangdong University of Technology, China (e-mail: lijian@gdut.edu.cn). 
\IEEEcompsocthanksitem Qinglin Zhao and Jincheng Cai are with the School of Computer Science and Engineering, Macau University of Science and Technology, Macau, China (e-mail: qlzhao@must.edu.mo; jincheng.cai@outlook.com).
\IEEEcompsocthanksitem Shuoyao Wang is with the College of Electronics and Information Engineering, Shenzhen University, Shenzhen, Guangdong 518060, China (email:sywang@szu.edu.cn).
\IEEEcompsocthanksitem Zhifeng Hao is with the Department of Mathematics, Shantou University, Shantou 515063, China (e-mail: haozhifeng@stu.edu.cn).
\IEEEcompsocthanksitem Naiqi~Wu is with the Macao Institute of Systems Engineering and the Collaborative Laboratory for Intelligent Science and Systems, Macau University of Science and Technology, Macao, and also the State Key Laboratory of Precision Electronic Manufacturing Technology and Equipment, Guangdong University of Technology, Guangzhou 510006, China (e-mail: nqwu@must.edu.mo).
}
}
\markboth{Preprint}{Fresh Keys, Stale Secrets: Numeric Leakage across Linked LLM Prompts}

\maketitle

\begin{abstract}
Small businesses repeatedly ask cloud language models about the same confidential prices, unit costs and margins. Numeric prompt disguises for such numbers are evaluated one prompt at a time, but a provider sees a firm's history, in which each fresh secret gives a new, independent view of the same numbers. Since real costs are not public, we measure on synthetic firms with known values how much an honest-but-curious provider recovers when it knows the disguise algorithm, holds sector-level price and margin statistics, and links a firm's queries. In one prompt, per-field scaling, a fixed-axis instantiation of Zhang et al.'s order-preserving substitution and Pr$\varepsilon\varepsilon$mpt's released sanitizer keep the attacker's margin-hit advantage over a prior-only guess between $-0.6$ and $+4.4$ percentage points. After 16 linked queries with fresh secrets, every disguise we test (Pr$\varepsilon\varepsilon$mpt at $\varepsilon\ge0.5$) has a target recovered significantly more often: under the order-preserving substitution, margin recovery rises from 25.0\% to 59.4\%. The same attacker recovers 29.4 percentage points more margins with fresh maps than with one reused map. Theory explains the contrast: one reused secret bounds all queries by a single full-precision view of the fields shown, whereas fresh scaling keys are identifiable. Except for Pr$\varepsilon\varepsilon$mpt at $\varepsilon=0.5$, the gains persist when the attacker's prior comes from another population model and when values drift by up to 5\% between queries; on our firms, whose prose cues are weak, reading the prose with a language model adds at most 2.5 points. Code and data: \url{https://github.com/Jamli58/fresh-keys-stale-secrets}.
\end{abstract}


\section{Introduction}
\label{sec:intro}

Small firms increasingly ask cloud language models for help with pricing and cost decisions. A bakery owner who asks how to price a wholesale order of croissants types in the price and the unit cost; a machine shop that asks whether to switch suppliers gives its current part cost. Prices, unit costs and the gross margins they imply can be trade secrets, whose disclosure to generative-AI services carries legal risk~\cite{hrdy2025trade}. They are not personal identifiers, so redacting personal data does not conceal them. And a firm rarely asks only once: it returns to the same numbers with a new quote request this week and a margin review the next.

Numeric prompt disguises aim to let a firm use such a model without exposing these numbers. Pr$\varepsilon\varepsilon$mpt~\cite{chowdhury2025preempt} replaces value tokens with noise calibrated by metric differential privacy (metric DP). Zhang et al.~\cite{zhang2026collaborative} replace numbers through an order-preserving secret map, let the cloud write code, and re-run that code locally on the true values. Earlier, secure outsourcing disguised inputs with secret invertible transformations and inverted the result~\cite{atallah2002secure,wang2011secure}. These designs are evaluated one prompt at a time, by a formal guarantee for one release or by judging what one disguised input retains. A provider, however, does not see one prompt. It sees an account's history, and for a small firm many of its prompts concern the same unchanged numbers. If leakage grows with this history, a disguise that passes a per-prompt evaluation can still expose a firm's margins after a few more questions about the same numbers.

Repeated releases of an unchanged value are a known hazard for randomized mechanisms~\cite{dinur2003revealing}. Local-privacy telemetry answers this hazard by reusing one randomized report per value~\cite{erlingsson2014rappor,ding2017telemetry}. Composition theorems bound the total leakage of mechanisms that carry a privacy budget~\cite{dwork2014algorithmic}, and reconstruction attacks on encrypted databases show that leakage under one fixed key accumulates over queries~\cite{kellaris2016generic,lacharite2018improved,grubbs2019learning}. Most numeric prompt disguises, however, have no privacy budget that bounds a history, and a stateless prompt sanitizer does the opposite of telemetry: it draws new secret randomness for every query, which looks safer because no two queries share a key. What is missing is a measurement of what a provider recovers from such a history. We ask three questions. Does per-prompt evaluation understate what a provider recovers from a firm's linked queries (Q1)? Does it matter whether the secret is drawn fresh for each query or reused, and why (Q2)? And do the answers depend on the attacker, on the synthetic population, and on numbers staying fixed across queries (Q3)?

Our starting point is that a fresh secret gives the provider a new, independent view of the same numbers. Under fresh scaling keys each query shows a new multiple of the prices; under a fresh order-preserving map on a fixed axis it shows a new noisy reading of where each value lies on that axis; under fresh noise it shows a new noisy copy of each value. Combining such views can locate the values, much as averaging repeated measurements reduces noise. A reused secret, in contrast, repeats one view, so under conditions we make precise a history discloses no more than one full-precision view of all the fields it shows. There is one exception: if every view depends on the values only through an invariant, such as their order, fresh secrets reveal that invariant and nothing more. Whether a disguise is safe for repeatedly queried numbers is therefore a property of the history it produces, and for the disguises we study, whether the secret is fresh or reused largely determines what that history discloses.

\begin{figure}[t]
  \centering
  \includegraphics[width=0.75\columnwidth]{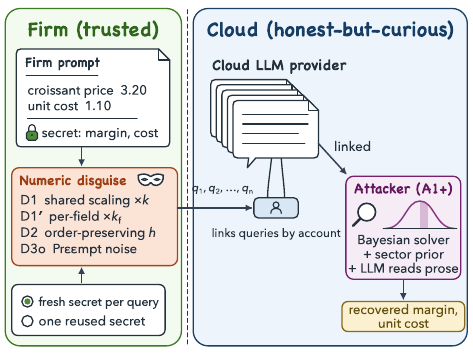}
  \Description{Threat model diagram. In the trusted firm region (left), a firm prompt with a croissant price and unit cost and a secret regime (fresh secret per query or one reused secret) feed a numeric disguise (D1, D1-prime, D2, D3o). Only the disguised queries q1 to qn cross the dashed trust boundary into the honest-but-curious cloud region (right), where the provider links them by account and passes them to the attacker A1+ (Bayesian solver with a sector prior, optionally aided by a language model reading the prose), which outputs the recovered margin and unit cost.}
  \caption{Threat model. Before a prompt leaves the firm, its numbers are disguised with a fresh or a reused secret; the provider links a firm's disguised queries and estimates the confidential margins and unit costs. Values shown are illustrative.}
  \label{fig:pipeline}
\end{figure}

We answer the three questions in turn. For Q1 we need ground truth, but real firms' unit costs and margins are not public, so real prompts cannot be scored. We therefore build counterfactual business prompts whose sector margins are anchored to public aggregates, the U.S. Internal Revenue Service (IRS) Statistics of Income and industry margin tables, and whose prose carries realistic cues such as the city and the market positioning. We implement four disguises that cover the three designs above: shared scaling, per-field scaling, an order-preserving substitution on a fixed public axis that instantiates Zhang et al.'s description, and Pr$\varepsilon\varepsilon$mpt's released money sanitizer. We attack them with a Bayesian attacker that knows each algorithm and holds a sector-level prior table, after one and after up to 16 linked queries (\cref{fig:pipeline}). For Q2, the comparison needs an attacker that can exploit a reused secret. Under reuse, repeated queries show identical values, so an attacker that reads one price--cost pair at a time sees nothing new and would make reuse look safe by construction. We therefore apply one joint attacker, which models all values shown under a map, to both regimes, and prove when reuse caps the history and when fresh secrets do not. For Q3, every reported rate is achieved by a concrete attacker, so a weak attacker could understate leakage in both regimes. We therefore test stronger numeric attackers, a language model that reads the prose, a fresh draw of firms, shifted populations, an attacker prior from another population model, and values that drift between queries.

\begin{figure}[t]
  \centering
  \includegraphics[width=0.75\columnwidth]{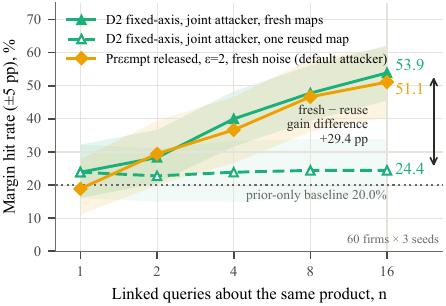}
  \Description{Line chart of margin hit rate against the number of linked queries (1 to 16). Under D2, the joint attacker with fresh maps rises from 24 to 54 percent and with one reused map stays between 23 and 24 percent; D3o at epsilon 2 rises from 19 to 51 percent. A double arrow at 16 queries marks the fresh-minus-reuse difference, and a dotted line marks the 20 percent prior-only baseline.}
  \caption{Margin recovery as linked queries accumulate (hit: error $\le5$\,pp). Both \Dtwo curves use the same joint attacker, with fresh maps and with one reused map; \Dthreeo uses the default attacker. Bands: firm-clustered 95\% bootstrap CIs; 60 test firms $\times$ 3 seeds.}
  \label{fig:hero}
\end{figure}

\noindent Our contributions are as follows.
\begin{enumerate}
\item \textbf{Ground-truth measurement of history leakage (Q1).} Recovery is scored against ground truth, and every departure of the four disguises from their original descriptions is stated. In one prompt on our synthetic firms, three of the four disguises keep the attacker's margin-hit advantage over a prior-only guess between $-0.6$ and $+4.4$\pp; only shared scaling exposes every margin. After 16 linked queries with fresh secrets, every disguise has a target recovered significantly more often (Pr$\varepsilon\varepsilon$mpt at $\varepsilon\ge0.5$); under the order-preserving substitution, margin recovery rises from 25.0\% to 59.4\%.
\item \textbf{Fresh versus reused secrets, measured and explained (Q2).} Applied to both regimes, one joint attacker gains 29.4 [18.3, 40.6]\pp more in margin hits with fresh maps than with one reused map (\cref{fig:hero}). Standard tools explain the contrast: by the data-processing principle, a reused secret bounds all queries of a field-local disguise with fixed field sets by one full-precision view of the fields shown, while fresh scaling keys are identifiable, and for log-uniform keys the feasible log-interval has expected width $2\log(b/a)/(n{+}1)$ after $n$ exact observations. A diagnostic result locates the leak of order-preserving substitution: a map that depends on ranks alone reveals only the order, so the leakage we measure comes from intervals fixed on a public axis.
\item \textbf{Boundaries of the result, and three checks (Q3).} The gains replicate on a fresh draw and on all 250 test firms, and except for Pr$\varepsilon\varepsilon$mpt at $\varepsilon=0.5$ they persist under population shifts, an attacker prior from another population model and value drift of up to 5\% between queries. Stronger likelihoods raise recovery, and on our firms, whose prose cues carry little price information, reading the prose with a language model adds at most 2.5\pp. From these results we derive three checks for disguise designers: evaluate a disguise against the history it produces, choose between fresh and reused secrets deliberately, and test whether its outputs depend on the values beyond an invariant.
\end{enumerate}

The rest of the paper is organized as follows. \Cref{sec:related} reviews related work. \Cref{sec:background} defines the mechanisms, the threat model and the problem, and \cref{sec:framework} the measurement framework. \Cref{sec:single,sec:linked} answer Q1, \cref{sec:regimes} answers Q2, and \cref{sec:bounds} answers Q3. \Cref{sec:proofs} gives the proofs, \cref{sec:discussion} discusses implications and limitations, and \cref{sec:conclusion} concludes. Sections, tables and figures numbered S1, S2, \ldots\ are in the supplemental material, which follows the conclusion in this preprint.

\section{Related Work}
\label{sec:related}

We review prior work on protecting and attacking numeric values in the light of one question: is leakage assessed for a single release, or across repeated releases of the same values? \Cref{tab:related} summarizes the closest works.

\begin{table*}[h!]
\centering
\footnotesize
\caption{Closest Prior Work and the Dimensions Measured Here}
\label{tab:related}
\begin{tabular}{@{}>{\raggedright\arraybackslash}p{1.8cm}>{\raggedright\arraybackslash}p{3.6cm}>{\raggedright\arraybackslash}p{4.6cm}>{\raggedright\arraybackslash}p{2.8cm}>{\raggedright\arraybackslash}p{3.6cm}@{}}
\toprule
Ref. & Target & Privacy assessment & Auxiliary information & Repeated releases of a secret \\
\midrule
\cite{chowdhury2025preempt} & identifiers and value tokens & metric DP per sanitization; input-dependent domain in released code & -- & stateless; linkage not analysed \\
\cite{zhang2026collaborative} & numbers and topic of financial text & language-model judge of retained information & -- & not discussed \\
\cite{siyan2025papillon} & personal information in queries & language-model judge of retained private data & -- & -- \\
\cite{staab2024beyond} & personal attributes in free text & attribute inference by language models & model world knowledge & -- \\
\cite{naveed2015inference,grubbs2017leakage} & property-preserving encrypted columns & plaintext recovery & public datasets & static database \\
\cite{kellaris2016generic,lacharite2018improved,grubbs2019learning} & values in encrypted databases & reconstruction from range-query leakage & query distribution & many queries under one fixed key \\
\cite{erlingsson2014rappor,ding2017telemetry} & client values in telemetry & local DP over repeated collection & -- & memoization bounds longitudinal leakage \\
\textbf{Ours} & margins, unit costs in business prompts & recovery rate against ground truth & sector priors, mechanism & 1--16 linked queries, fresh vs reused \\
\bottomrule
\end{tabular}
\par\vspace{2pt}\parbox{\textwidth}{\footnotesize\raggedright --: not evaluated or not discussed in that work.}
\end{table*}

\paragraph{Protecting numbers in prompts} \preempt~\cite{chowdhury2025preempt} encrypts identifiers with format-preserving encryption and sanitizes value tokens with metric differential privacy. Its paper gives a formal guarantee for each sanitization of the published construction, which the released money functions do not inherit across domain boundaries (\cref{sec:background}). Zhang et al.~\cite{zhang2026collaborative} protect numbers in financial reasoning through topic rewriting, order-preserving substitution and local re-execution of cloud-written code, and ask a language model whether the synthesized input retains original information. Other systems delegate between local and cloud models~\cite{siyan2025papillon}, perturb a prompt's embeddings with local differential privacy~\cite{du2023dpforward}, or anonymize text against an inferring adversary~\cite{staab2025anonymizers}. Token-level text sanitizers replace words by substitutes drawn under metric local differential privacy~\cite{yue2021santext,chen2023custext}, DP-Prompt paraphrases a document under local differential privacy~\cite{utpala2023dpprompt}, and InferDPT perturbs a prompt before closed-box inference and restores the answer locally~\cite{tong2025inferdpt}; an empirical comparison of prompt-privacy techniques measures leakage on a ground-truth-labelled benchmark~\cite{agyemang2026llmredactor}. These works assess privacy one sanitization or one disguised input at a time.

\paragraph{Disguise-based outsourcing and its attacks} Disguising inputs with secret linear or matrix transforms so that an untrusted server computes on them dates back to Atallah et al.~\cite{atallah2002secure} and was refined for linear programming~\cite{wang2011secure}. Shared and per-field scaling are the simplest members of this family. Structure-preserving disguises fall to attackers that combine the structure with auxiliary information: multiplicative perturbation is open to recovery by independent component analysis~\cite{liu2006random}, distance-preserving perturbation can be reconstructed from a few known inputs~\cite{giannella2013breaching}, and property-preserving encryption leaks plaintexts to attackers with auxiliary data~\cite{naveed2015inference,grubbs2017leakage}. Our attacker belongs to this tradition: it combines the mechanism's structure with a prior table. These attacks, however, target one static collection of disguised values. A prompt holds only a few values, and the larger gains come from repeated transformations of the same values across prompts, a setting the outsourcing literature rarely models.

\paragraph{Leakage across repeated releases} Closest to our question is work in which leakage grows as releases or queries accumulate. Randomized reports of an unchanged value accumulate evidence, so local differential privacy for telemetry memoizes a randomized response per value, permanently in RAPPOR~\cite{erlingsson2014rappor} and for repeated numeric reports in Ding et al.~\cite{ding2017telemetry}. Differential privacy composes, so $n$ releases under $\varepsilon$ each are $n\varepsilon$-private~\cite{dwork2014algorithmic}, and the same holds for metric differential privacy under a fixed metric~\cite{chatzikokolakis2013broadening}. Without such a budget, repeated releases are dangerous: many answers about the same data with small noise allow its reconstruction~\cite{dinur2003revealing}, and independent releases combined disclose what none shows alone~\cite{ganta2008composition}. On encrypted databases, reconstruction attacks show that the access pattern or result volume of range queries under one fixed key exposes the stored values as queries accumulate~\cite{kellaris2016generic,lacharite2018improved,grubbs2019learning}.  Prompt disguises differ from these lines. The scaling and order-preserving disguises carry no formal budget, and released \preempt, with its input-dependent domain, falls outside a global metric-DP accounting (\cref{sec:background}), so composition does not bound their history. Unlike the database attacks, a prompt disguise may draw a fresh secret for every query, what leaks are the transformed values themselves rather than access patterns, and the attacker holds only sector-level quantiles. What these results leave open is how much a provider recovers from the histories that concrete prompt disguises produce, and whether reusing the secret helps. Our reuse ceiling (\cref{thm:reuse}) applies the data-processing principle to field-local disguises, the counterpart of memoization, and \cref{thm:rate} gives the shrinkage of the feasible set for fresh scaling keys.

\paragraph{Summary} Language models also infer personal attributes from text~\cite{staab2024beyond}, which \cref{sec:llm} tests for business numbers. What is missing is a measurement of how numeric prompt disguises leak across a firm's linked queries, and of whether a fresh or a reused secret makes this leakage worse. This paper provides that measurement and a theory of the contrast.

\section{Mechanisms, Threat Model and Problem Statement}
\label{sec:background}

This section fixes what is attacked, what the attacker knows, and what we measure. The four mechanisms cover the three designs of \cref{sec:intro}; the scaling and order-preserving mechanisms can run with a fresh or a reused secret, the variable this paper studies; released \preempt draws fresh noise for every release.

\subsection{Numeric disguises for prompts}
A prompt contains typed numeric fields $x = (x_f)_{f\in\Fset}$ (prices, unit costs, overhead, volumes, years). A disguise draws a secret and replaces each field by a shown value; the client keeps whatever it needs to restore the answer. We study four mechanisms (\cref{tab:mech}); ``fresh'' draws a new secret for every query and ``reuse'' keeps one secret for a firm.

\begin{table}[t]
\centering
\footnotesize
\caption{Evaluated Mechanisms}
\label{tab:mech}
\setlength{\tabcolsep}{2pt}
\begin{tabular}{@{}ll>{\raggedright\arraybackslash}p{2.35cm}@{}}
\toprule
Mechanism & Secret & Visible in one prompt \\
\midrule
\Done shared scaling & $k\sim\mathrm{LogUniform}[0.1,10]$ & money ratios, hence margins \\
\Donep per-field scaling & $k_f\sim\mathrm{LogUniform}[0.1,10]$ & no shared-key ratio \\
\Dtwo fixed-axis order-pres. & increasing map $h$ & order; axis-dependent law \\
\Dthreeo \preempt (released) & noise, $\varepsilon$ per value & domain-dependent support \\
\bottomrule
\end{tabular}
\end{table}

\paragraph{Shared and per-field scaling (\Done, \Donep)} Secure outsourcing disguises inputs by secret linear maps and inverts results~\cite{atallah2002secure,wang2011secure}. \Done multiplies every money field by one key $k\sim\mathrm{LogUniform}[0.1,10]$ and shows whole cents. \Donep, our control, draws an independent key per field, which removes the ratio invariant.

\paragraph{Fixed-axis order-preserving substitution (\Dtwo)} \Dtwo instantiates the order-preserving substitution of Zhang et al.~\cite{zhang2026collaborative}. Their system shifts year-like integers in $[1990,2030]$ by a shared offset, leaves designated special numbers (such as 28--31) unchanged, sorts the remaining numbers into intervals and maps them to sorted random targets while preserving order, rewrites the topic with a local model, and re-runs cloud-written code locally on the true values. We could not locate a public implementation. \Dtwo keeps their year rule, the special numbers 28--31 and their order-preserving rule, omits the topic shift, and fixes what the description leaves open: the intervals are 64 equal cells of a public signed-log axis $t=\mathrm{sign}(x)\log(1+|x|)$, and a keyed increasing map $h$ interpolates in log space between random endpoint targets in $[1,10^6]$ (sampling law in \cref{app:mech}); the year offset is drawn from $[-10,10]\setminus\{0\}$.

Because the axis is public, the law of $h(x)$ depends on where $x$ lies and not only on its rank; a substitution whose output depends on ranks alone has no such channel (\cref{prop:rank}). \Cref{tab:d2fid} separates the inherited rules from our choices. \Dtwo is therefore a controlled study of a public-axis order-preserving channel rather than a reproduction of Zhang et al.'s system, and all \Dtwo results concern this construction.

\begin{table}[t]
\centering
\footnotesize
\caption{\Dtwo Relative to Zhang et al.'s Description~\cite{zhang2026collaborative}}
\label{tab:d2fid}
\setlength{\tabcolsep}{3pt}
\begin{tabular}{@{}>{\raggedright\arraybackslash}p{0.29\columnwidth}>{\raggedright\arraybackslash}p{0.10\columnwidth}>{\raggedright\arraybackslash}p{0.32\columnwidth}>{\raggedright\arraybackslash}p{0.19\columnwidth}@{}}
\toprule
Component & Source & In \Dtwo & Varied \\
\midrule
Year offset, $[1990,2030]$ & \cite{zhang2026collaborative} & offset range $[-10,10]\setminus\{0\}$ (ours) & -- \\
Special numbers unchanged & \cite{zhang2026collaborative} & 28--31 only (1 and 12 mapped) & -- \\
Sorted intervals and targets & \cite{zhang2026collaborative} & as described & -- \\
Interval layout & ours & 64 equal cells of public signed-log axis & 16, 256 cells \\
Endpoint targets & ours & in $[1,10^6]$, uniform cuts, log-linear interpolation & Dirichlet(0.1), (10); $[1,10^{12}]$ \\
Display precision & ours & fewest digits ${\ge}6$ keeping inputs distinct & -- \\
Topic shift (local model) & \cite{zhang2026collaborative} & omitted & -- \\
Local re-run of cloud code & \cite{zhang2026collaborative} & not modelled (utility only) & -- \\
Alternative: map of ranks only & -- & not evaluated (\cref{prop:rank}) & -- \\
\bottomrule
\end{tabular}
\par\vspace{2pt}\parbox{\columnwidth}{\footnotesize\raggedright ``Varied'': alternatives evaluated in \cref{S-tab:d2law}.}
\end{table}

\paragraph{Metric-DP value tokens (\Dthreeo, \Dthree)} \preempt~\cite{chowdhury2025preempt} sanitizes value tokens with metric differential privacy. \Dthreeo is our implementation of the money-sanitizer functions in \preempt's released code~\cite{preemptcode} (\texttt{encypt\_money} and \texttt{M\_epsilon}), applied to typed money fields with the full $\varepsilon$ per value; counts and years are unchanged. It truncates a value to an integer $v$, chooses a domain $[\ell,u]$ by the magnitude of $v$, rescales $v'=100(v-\ell)/(u-\ell)$, samples a bin $s\in\{0,\dots,99\}$ with probability $\propto\exp(-|v'-s|\varepsilon/2)$ and shows the integer part of $\ell+s(u-\ell)/100$ (domains in \cref{S-sec:mech}). Within one domain these probabilities satisfy the metric-DP inequality $\Pr[s\mid v_1]\le e^{\varepsilon|v'_1-v'_2|}\Pr[s\mid v_2]$~\cite{chatzikokolakis2013broadening,andres2013geo}. The domain is itself chosen from the private value, however, so the output support depends on the input: 2 is a possible output for $v=99$ but not for $v=100$. The released wrapper therefore has no finite $\varepsilon$ across domain boundaries, and our likelihood includes this support information. As a contrast we also report \Dthree, our variant of the paper's description: each of the $N_{\mathrm{v}}$ value tokens of a prompt receives $\varepsilon/N_{\mathrm{v}}$ and is replaced by a truncated-Laplace draw on a fixed domain (\cref{S-sec:mech}), so equal nominal $\varepsilon$ is not an equal budget for \Dthreeo and \Dthree.

\subsection{Threat model}
The adversary is an honest-but-curious cloud provider. It sees every disguised prompt and knows each mechanism and its public parameters, but not the secrets (keys, maps, noise). It holds an \emph{assumed} prior table: the P10/Q25/median/Q75/P90 of price, unit cost and margin per sector and product class. It is also told each product's sector and product class, which product names usually indicate (a jacket in an apparel prompt); the language-model attacker of \cref{sec:llm} infers both from the prose instead. Its targets are each product's gross margin $m=(p-c)/p$ and unit cost $c$, and for scaling also the key.

In the linked-query setting (\cref{sec:linked}) the provider additionally links a firm's queries by account and knows which shown field corresponds to which product across queries, and the firm's target values are unchanged across queries (\cref{sec:mismatch} relaxes this last assumption); we evaluate this setting with the numeric attackers only. This is a worst case for linkage, and the natural one for a provider serving a logged-in business account.

\subsection{Problem statement}
\label{sec:problem}
A margin estimate is a \emph{hit} if its absolute error is at most 5 percentage points (pp); a unit-cost estimate is a hit if its relative error is at most 10\%. These are tolerance-based guessing criteria in the sense of gain-function leakage measures~\cite{alvim2012gain}. Let $H_n$ be an attacker's hit rate for a target after $n$ linked queries of a firm, under the fresh or the reused regime. A per-prompt evaluation reports $H_1$. We measure the history gain $H_{16}-H_1$ of each mechanism with fresh secrets (Q1), the fresh-minus-reuse difference, that is, the fresh history gain minus the reused one under the same attacker (Q2), and how both change with the attacker and the population (Q3).

The \emph{advantage} of an attack is the paired difference of its hit rate to a stated baseline: the prior-only attacker \Azero for single prompts, and the default attacker \Aplus for the language-model attackers (\cref{sec:llm}). Hit rates average over rows, so single-query rates weight product observations equally and linked-query rates weight firms equally. Every hit rate is achieved by a concrete attacker and is therefore a lower bound on what a provider with the same information can learn; an advantage compares two such attackers and is not a bound.
\section{Measurement Framework}
\label{sec:framework}

This section builds the tools that Q1--Q3 share: prompts whose true values are known, the attackers, the evaluation protocol, and checks of the implementation.

\subsection{Counterfactual business prompts}
We generate 300 counterfactual firms (50 for development, 250 for test) in six sectors: bakery/caf\'e, machine shop, apparel e-commerce, IT consulting, construction subcontractor and specialty retail. Each firm sells three to eight products. Prices follow a product-class median times a city price level, a firm brand premium and heavy-tailed noise; margins follow a sector median shifted by a latent firm efficiency, and each firm is drawn as extreme with probability 0.2. Sector margin medians are anchored to public aggregates: one minus the ratio of cost of goods sold to business receipts in the IRS Statistics of Income corporation tables for tax year 2022~\cite{irs2022soi} for five sectors, and the January 2026 Computer Services gross margin~\cite{damodaran2026margins} for IT consulting, both modeling choices since an aggregate ratio is not a median firm's margin. Generated sector medians are within 0.02 of these anchors, and 22.7\% of firms fall outside their sector's P10--P90 margin band. Prices, their dispersion within a class, volumes and overhead are documented assumptions (\cref{S-sec:data}).

Each firm's prompts name one of 60 US metropolitan areas sampled across the 2024 Regional Price Parities~\cite{bea2024rpp} and a positioning (premium, standard or discount). These cues are informative but weak: the correlation between the log price parity and a firm's log price residual is 0.26. The attacker's prior table is computed from a separate sample of 2{,}000 firms drawn from the same generator, an independent draw rather than a different population.

Five task templates (quote pricing, margin review, break-even, monthly sales summary, supplier comparison) turn a firm into a prompt; every firm contributes two tasks. The test split has 500 prompts with 700 product observations (450 distinct products); with three defense seeds this gives 2{,}100 rows per attacker and mechanism. Every number in a prompt is a typed field, and a test checks that no number appears in the prose outside a field. Only the numbers change under a disguise; the prose, including the city and positioning cues, stays identical. In the margin-review prompt of the first test firm, for example, a jacket priced 104.03 with unit cost 90.01 is shown as 195.47 and 169.13 under \Done (same margin) and as 298 and 81 under \Dthreeo at $\varepsilon=1$ (all four in \cref{S-sec:data}).

\subsection{Attackers}
\label{sec:attackers}
The default attacker \Aplus is used for every mechanism; the language-model attackers are described in \cref{sec:llm}.
\paragraph{Prior-only (\Azero)} Predicts the class median margin and unit cost from the prior table, given the true sector and class.
\paragraph{Classical Bayesian (\Aplus)} The default attacker knows the mechanism and is told the true sector and product class, as is \Azero. Its prior draws log price and logit margin independently from Student-$t$ distributions with 3 degrees of freedom, centred at the class medians of the prior table, with scales that match the table's interquartile ranges, multiplied by 0.75; shape and width were chosen on the development split from six candidates and then frozen for every mechanism and attacker. It reports the posterior median of each product's margin and unit cost, which minimizes absolute error rather than maximizing the hit rate; \cref{sec:single} also evaluates hit-optimal decisions. The likelihood depends on the mechanism:
\begin{itemize}
\item \Done: the margin follows from the scale invariant, and the key from a grid posterior that combines the log-uniform key prior with each product's price prior.
\item \Donep, \Dthreeo, \Dthree: importance sampling with 4{,}000 prior draws, weighted by the exact likelihood of the shown values for \Donep (accounting for whole-cent rounding) and \Dthreeo, and by the truncated-Laplace density at the shown value for \Dthree.
\item \Dtwo: each (price, cost) pair is scored jointly with a bivariate Gaussian approximation of $(\log h(p),\log h(c))$, whose moments are simulated over 16{,}000 secret maps.
\end{itemize}
\paragraph{Linked queries} The multi-query attacker extends \Aplus to a firm's sequence of prompts and targets product~0, which every template shows. Under \Done it uses the key ratios between queries and the public key range; under the other mechanisms it sums product~0's log-likelihood over queries, counting deterministic repeats under a reused secret once. With one query it reproduces \Aplus exactly.
\paragraph{Attackers for \Dtwo} \Dtwo is attacked by four attackers that differ in which shown values they model and how they approximate the likelihood, all with the same prior, oracle and decision rule; the three stronger ones serve \cref{sec:regimes,sec:bounds}. The default \Aplus, the \emph{pair-only attacker}, scores product~0's (price, cost) pair with the Gaussian approximation above. A \emph{reference likelihood} replaces that approximation with a kernel density over 20{,}000 simulated maps (bandwidth 0.03 in log units). A \emph{joint shared-map attacker} models the prices and costs of all products shown under one map, and integrates the other products' unchanged values over their class priors. \emph{A-exact} is a particle approximation to the posterior under the exact display model: it samples maps from the public law, inverts every shown value and integrates the display rounding, and is used for single queries and reused maps (\cref{app:attack}).

\subsection{Evaluation protocol}
\label{sec:protocol}
Attacker choices were made on the development split. The classical attackers were evaluated on the test split several times during development, including after the generator's margins were replaced by the cited anchors. To guard against adaptive overfitting, we confirm the linked-query results on the first 60 test firms of a fresh draw of 300 firms (new generator seed, same prior table, frozen attacker; \cref{sec:robust}). The language-model attackers were evaluated on the test split once, with endpoints and a success criterion fixed beforehand; a third prespecified endpoint (C1c), defined under Zhang et al.'s topic shift, was not run because \Dtwo omits the topic shift.

All intervals over firms are paired percentile bootstrap 95\% confidence intervals (CIs) that resample firms with all their rows (products and defense seeds), with 10{,}000 resamples (2{,}000 for the oracle-cue control of \cref{sec:llm}). The headline linked-query claim is checked with a Holm correction (\cref{sec:linked}); the other comparisons carry no multiplicity adjustment and are reported with their own intervals. Code, all result files (including the language-model outputs, which allow a replay without model calls), the prespecified plans and an evidence index are public at \url{https://github.com/Jamli58/fresh-keys-stale-secrets}.

\subsection{Implementation checks and calibration}
\Dthreeo's likelihood equals the probability mass function of the released \texttt{M\_epsilon} for every output and every tested value and $\varepsilon$ (checked against a verbatim copy of the released function). Our \Dtwo reconstructs the original execution on all 1{,}108 arithmetic FinQA test items~\cite{chen2021finqa} when the gold program stands in for the cloud's code, with no display collisions, a check of reconstruction correctness only. \Done is shown in whole cents, so inversion from shown values is exact only up to rounding.
\paragraph{Self-consistency} This check calibrates what a small single-prompt advantage means. Products drawn from the attacker's own prior and disguised are recovered better than the prior median under \Done (+62.0 [59.9, 64.1]\pp margin hit) and \Dthree, but only by +0.8 [$-$0.4, 1.9]\pp under \Donep and +1.2 [$-$1.3, 3.6]\pp under \Dtwo, so a small single-prompt advantage under these two is expected even when the prior is correct.

\subsection{Details of the \Dtwo attackers}
\label{app:attack}
\label{app:mech}
This subsection specifies the map law of \Dtwo and the two stronger attackers that \cref{sec:regimes} uses to compare fresh and reused maps; the results can be read without these details.

\paragraph{\Dtwo} The signed-log axis $t=\mathrm{sign}(x)\log(1+|x|)$ on $[-\log(1+10^{12}),\log(1+10^{12})]$ is cut into 64 equal intervals; their endpoints receive increasing log-targets from $0$ to $\log10^6$: each interval's log-gap is a floor of 0.01 times its width in $t$ (the minimum slope) plus a share of the remaining range, with shares given by uniform sorted cuts (Dirichlet(1); \cref{S-tab:d2law} varies this), and $h$ interpolates linearly in log-target space. Values are shown with the fewest significant digits ($\ge6$) that keep distinct inputs distinct. Integers in $[1990,2030]$ are shifted by a secret offset in $[-10,10]\setminus\{0\}$; integers 28--31 pass through.
\paragraph{Joint shared-map attacker} For the products shown under one map, the log targets of all their prices and costs are modelled as multivariate Gaussian with moments simulated from the map law, plus an observation smoothing of 0.03 in log units on each value (without it, values that share a map interval are almost collinear and the integral below cannot be resolved numerically). The other products' values are unchanged across queries, so one set of 400 draws $z_s$ of these values from their class priors is shared by all queries. The likelihood of product~0's candidate $\theta$ (its log price and log unit cost) is the average over the draws of $\prod_g q_g(Y_g\mid\theta,z_s)$. Here $q_g$ is the smoothed Gaussian density of the log values $Y_g$ shown under map $g$, with one factor per map: one per query under fresh maps, and one for the union of shown products under a reused map. A query that shows only product~0 is scored by the pair model. A unit test checks the Monte Carlo integral against two-dimensional quadrature.
\paragraph{Particle attacker for the exact display model (A-exact)} A-exact treats as unknowns the map's knot vector $\Lambda$ and the latent log-output $u_i$ of every shown mapped field. A field is consistent with its shown value exactly when $u_i$ lies in the log rounding cell $C_i$ of that value. With $\mathcal P$ the products and $\mathcal O$ the other mapped fields, the target is
\begin{equation*}
\pi(\Lambda,u)\propto p(\Lambda)\prod_{k\in\mathcal P}g_k J_k\prod_{i\in\mathcal O}\Bigl|\frac{\partial t_i}{\partial u_i}\Bigr|\prod_i\ind\{u_i\in C_i\}.
\end{equation*}
The four factors are the map law; the oracle class prior $g_k$ of each product's price and unit cost, with the Jacobian $J_k$ of the change of variables to $u$; a uniform prior on the axis for the other mapped fields, which leaves only their Jacobian; and the display constraint. The marginal of $\pi$ in the values is their posterior under the exact display model (derivation in \cref{S-sec:attackers}). The first query uses 400{,}000 maps, resampled to 20{,}000 particles; under a reused map each particle keeps its map, and after each query the particles are resampled and moved by 40 Metropolis--Hastings steps. Against exact rejection sampling on twelve small instances (six or eight intervals, one significant digit), the posterior quantiles (0.1, 0.5, 0.9) of product~0's margin agree within 0.012 and the Kolmogorov--Smirnov distance stays below the 1\% critical value for the attacker's effective sample size in every instance. Under fresh maps it would need, for every new map, the probability that the map rounds all previously pinned values to their shown digits; this weight degenerates as queries accumulate, so no fresh-map A-exact estimate is reported. Endpoints and numerical checks were fixed in a version-controlled plan before the main runs.

\section{Single Prompts: The Per-Prompt View}
\label{sec:single}

\begin{table*}[t]
\centering
\footnotesize
\caption{Single-Query Leakage on the Test Split}
\label{tab:single}
\resizebox{\textwidth}{!}{\begin{tabular}{lrrrrr}
\toprule
Mechanism & Margin hit (\%) & Cost hit (\%) & Margin MAE (pp) & $\Delta$ margin vs A0 (pp) & $\Delta$ cost vs A0 (pp) \\
\midrule
Prior only (A0) & 28.1 & 16.0 & 13.3 & -- & -- \\
\midrule
No disguise (D0) & 100.0 & 100.0 & 0.0 & +71.9 [+67.2, +76.4] & +84.0 [+80.2, +87.7] \\
Shared scaling (D1) & 100.0 & 28.6 & 0.026 & +71.9 [+67.2, +76.4] & +12.6 [+7.1, +18.0] \\
Per-field scaling (D1$'$) & 28.4 & 19.4 & 12.7 & +0.2 [-1.5, +2.0] & +3.4 [+1.0, +5.7] \\
Fixed-axis order-pres.\ (D2) & 29.1 & 15.8 & 11.4 & +1.0 [-2.7, +4.5] & -0.2 [-3.0, +2.5] \\
Pr$\varepsilon\varepsilon$mpt released (D3o), $\varepsilon$=0.5 & 27.5 & 26.1 & 12.2 & -0.6 [-3.4, +2.0] & +10.1 [+6.1, +14.0] \\
Pr$\varepsilon\varepsilon$mpt released (D3o), $\varepsilon$=1.0 & 29.5 & 31.5 & 11.7 & +1.4 [-1.6, +4.3] & +15.5 [+11.1, +19.8] \\
Pr$\varepsilon\varepsilon$mpt released (D3o), $\varepsilon$=2.0 & 32.5 & 38.2 & 10.6 & +4.4 [+1.0, +7.8] & +22.2 [+17.5, +26.9] \\
\midrule
Pr$\varepsilon\varepsilon$mpt variant (D3), $\varepsilon$=0.5 & 45.1 & 45.5 & 9.6 & +17.0 [+12.5, +21.6] & +29.5 [+24.4, +34.7] \\
Pr$\varepsilon\varepsilon$mpt variant (D3), $\varepsilon$=1.0 & 51.8 & 53.7 & 8.2 & +23.6 [+18.7, +28.6] & +37.7 [+32.2, +43.2] \\
Pr$\varepsilon\varepsilon$mpt variant (D3), $\varepsilon$=2.0 & 59.2 & 61.6 & 6.7 & +31.1 [+25.9, +36.3] & +45.6 [+39.9, +51.3] \\
\bottomrule
\end{tabular}}
\par\vspace{2pt}\parbox{\textwidth}{\footnotesize\raggedright 2{,}100 rows (700 product observations $\times$ 3 defense seeds); attacker \Aplus. Hit: margin error $\le5$\,pp, unit-cost error $\le10\%$. $\Delta$: paired difference to \Azero with firm-clustered 95\% bootstrap CI. \Dthree: our continuous fixed-domain variant of \preempt's paper description with $\varepsilon/N_{\mathrm{v}}$ per value token, so its $\varepsilon$ does not match \Dthreeo's.}
\end{table*}

\begin{figure}[t]
  \centering
  \includegraphics[width=0.92\columnwidth]{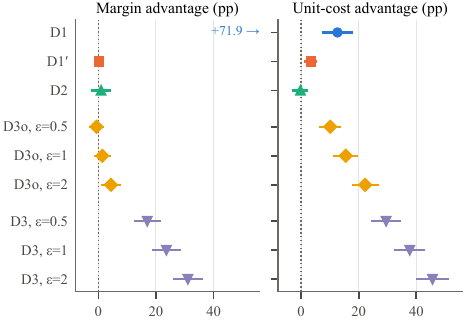}
  \Description{Dot plot of the single-query advantage of the default attacker over the prior-only attacker, for margin hits (left) and unit-cost hits (right), with 95 percent intervals, for D1, D1-prime, D2, D3o at three epsilons and the paper-reading variant D3 at three epsilons. Margin advantages of D1-prime, D2 and D3o lie near zero; D1 reveals margins; D3o shows unit-cost advantages; D3 shows large advantages on both.}
  \caption{In one prompt, the default attacker recovers margins at about the prior-only rate except under shared scaling, while released \preempt leaks unit costs. Single-query advantage of \Aplus over the prior-only attacker \Azero on the test split (2{,}100 rows per mechanism; bars: firm-clustered 95\% bootstrap CIs). \Done's margin advantage, +71.9\,pp, is off scale. \Dthree is our variant built from \preempt's paper description; values in \cref{tab:single}.}
  \label{fig:single}
\end{figure}

We first ask what a per-prompt evaluation concludes, the starting point of Q1. Apart from shared scaling, the disguises appear to protect margins: in one prompt the default attacker recovers them at about the prior-only rate. \Cref{tab:single} and \cref{fig:single} give the results. Without a disguise the default attacker \Aplus reads every value; the prior-only attacker \Azero, which knows the sector and class but ignores the numbers, hits 28.1\% of margins and 16.0\% of unit costs.

\paragraph{Shared scaling exposes every margin} A margin $(p-c)/p$ is invariant to a common factor, so \Done shows it exactly before display rounding; after whole-cent rounding all 2{,}100 margins are within $\pm5$\pp (mean absolute error, MAE, 0.026\pp). No attack is needed for this. The key still hides the absolute scale: unit-cost hits reach 28.6\%, an advantage of +12.6 [7.1, 18.0]\pp over \Azero.

\paragraph{Against three mechanisms, \Aplus recovers margins at about the prior-only rate} Independent per-field keys (\Donep), our order-preserving implementation (\Dtwo) and released \preempt (\Dthreeo) keep \Aplus close to \Azero on margins (\cref{fig:single,tab:single}): the advantages lie between $-0.6$ and $+4.4$\pp, and only \Dthreeo at $\varepsilon=2$ has an interval above zero (+4.4 [1.0, 7.8]\pp). The small advantage holds with hit-optimal decisions instead of posterior medians (at most +4.8\pp) and on a fresh draw of firms ($-0.9$ to 0.0\pp for \Donep, \Dtwo and \Dthreeo at $\varepsilon=1$). It does depend on how well the prior fits: on the 498 rows of atypical firms \Azero hits no margin and \Aplus at most 8.2\% (\cref{S-tab:strata}), and with margins 0.5 logit below the prior table the shown numbers add +9.7 [6.2, 13.3]\pp under \Dtwo (\cref{S-sec:single}).
\paragraph{Released \preempt leaks unit costs} Unit costs leak under \Dthreeo (+10.1, +15.5 and +22.2\pp, all intervals above zero), in part through its input-dependent support alone: when releases depend on the value almost only through the domain ($\varepsilon=10^{-6}$), the unit-cost advantage is still +4.8 [0.9, 8.6]\pp. Our variant built from \preempt's paper (\Dthree) leaks far more per prompt, +17.0 to +31.1\pp margin advantage.
\paragraph{Stronger attackers on \Dtwo} Is the small \Dtwo advantage only a weakness of the default attacker? The reference likelihood gives a margin advantage of only +1.8 [$-$1.0, 4.4]\pp, but attackers that model several values under one map recover more: on the 300 product-0 rows of the margin-review prompts, which show three products, A-exact gains +10.0 [2.0, 18.0]\pp in margin hits over \Azero, and +3.4 [$-$0.5, 7.2]\pp on all 1{,}500 product-0 rows (\cref{S-sec:single}). The exact within-interval relations thus already leak in one prompt when several values share the map.

Judged one prompt at a time, then, shared scaling exposes every margin, while under the other three mechanisms the default attacker's margin advantage is at most 4.4\pp; under \Dtwo the strongest attacker reaches +10.0\pp on prompts that show several products. \Cref{sec:linked} asks whether this judgment holds when a firm queries the same numbers again.

\section{Linked Queries: Leakage Grows with the History}
\label{sec:linked}

This section completes Q1: what happens when a firm asks about the same numbers 16 times and each query draws a fresh secret.

\subsection{Setup}
We take the first 60 test firms in generation order (ten per sector) and let each ask 16 questions that cycle through the five templates. Every template shows product~0's price and cost, so product~0 is the target. The templates differ in what else they show; margin review, for instance, adds the prices and costs of up to two more products (\cref{S-sec:data}). The threat model of \cref{sec:background} applies.  Under the \emph{fresh} regime every query draws a new secret (a new key, map or noise draw); under \emph{reuse} one secret serves the whole sequence (\Done, \Donep, \Dtwo). \Dthreeo, the money-noise branch of \preempt, has no reusable secret: each release draws fresh noise. The multi-query \Aplus of \cref{sec:framework} attacks the first $n\in\{1,2,4,8,16\}$ queries of the same disguised sequence, so the curves are paired across $n$. Each point pools 60 firms and three defense seeds (180 rows), so intervals span about $\pm10$\pp (\cref{sec:robust} repeats the comparison on all 250 test firms); the prior-only baseline for product~0 is 20.0\% for margins and 15.0\% for unit costs. \Cref{sec:regimes} compares the regimes on the same sequences.

\begin{figure*}[t]
  \centering
  \includegraphics[width=0.68\textwidth]{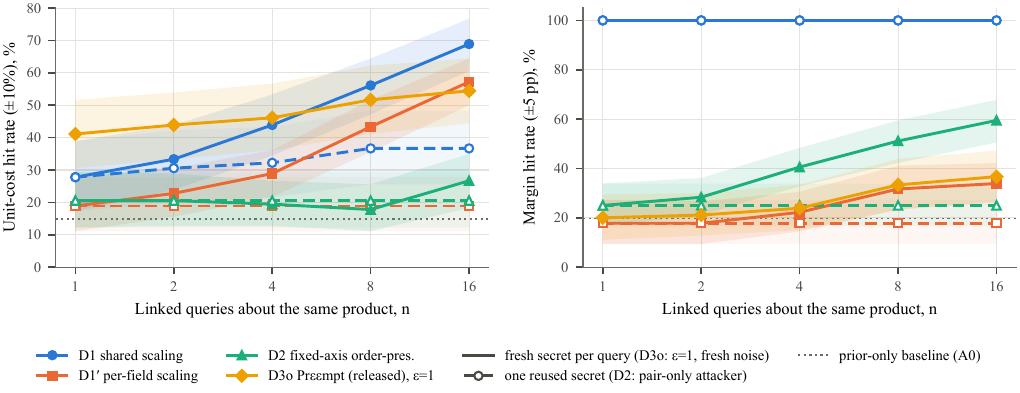}
  \Description{Two panels of hit rate against linked queries (1 to 16): unit-cost hits within 10 percent and margin hits within 5 percentage points, for D1, D1-prime, D2 with fresh and reused secrets and D3o at epsilon 1. With fresh secrets, unit-cost hits rise for D1, D1-prime and D3o, margin hits rise for D1-prime, D2 and D3o, D1 margins stay at 100 percent, and D2 unit-cost hits are non-monotone; with reused secrets the curves stay flat except a small, non-significant rise for D1 unit cost.}
  \caption{With a fresh secret per query, recovery grows with the number of linked queries; with one reused secret it stays near its single-query level. Recovery of product~0 by the default attacker \Aplus. Solid: a fresh secret per query (\Dthreeo: fresh noise at $\varepsilon{=}1$); dashed: one reused secret. \Done recovers 100\% of margins in both regimes (the lines coincide). The reused-map \Dtwo curve is from the pair-only attacker; \cref{tab:d2valid} compares the regimes under the joint attacker. Dotted: prior-only baseline (15.0\% cost, 20.0\% margin). 180 rows per $n$ (60 test firms $\times$ 3 seeds); bands: firm-clustered 95\% bootstrap CIs.}
  \label{fig:linked}
\end{figure*}

\subsection{Results}
\Cref{fig:linked} shows the result. With a fresh secret per query, every mechanism has a positive history gain $H_{16}-H_1$ for at least one target, also after a Holm correction over the 18 setting--target tests below (each setting's strongest target has adjusted $p\le0.0014$):
\begin{itemize}
\item \Done (shared scaling): unit-cost hits rise from 27.8\% to 68.9\% (+41.1 [30.6, 51.1]\pp); the median error of the first key, $|\log\hat k_1-\log k_1|$, falls from 0.203 to 0.066. Margins are recovered in every query.
\item \Donep (per-field scaling): unit-cost hits rise from 18.9\% to 57.2\% (+38.3 [28.3, 47.8]\pp) and margin hits from 17.8\% to 33.9\% (+16.1 [7.8, 23.9]\pp). Removing the shared key removes the single-query invariant but not the accumulation.
\item \Dtwo (order-preserving substitution): margin hits rise from 25.0\% to 59.4\% (+34.4 [23.9, 45.0]\pp). Its unit-cost curve is not monotone (20.6, 20.6, 19.4, 17.8, 26.7\%) and its gain, +6.1 [$-$3.3, 15.0]\pp, is not significant.
\item \Dthreeo (released \preempt): margin hits gain +8.3 [2.2, 15.0], +16.7 [8.3, 25.0] and +32.2 [23.3, 41.7]\pp at $\varepsilon=0.5,1,2$ (from 18.9\% to 51.1\% at $\varepsilon=2$), and unit-cost hits +14.4, +13.3 and +23.9\pp, all with intervals above zero. Domain selection alone does not produce this growth: when releases depend on the value almost only through its domain ($\varepsilon=10^{-6}$), the gains are +0.6 [0.0, 1.7]\pp for margins and +1.7 [0.0, 3.9]\pp for unit costs.
\item \Dthree (variant) gains +13.3 to +21.7\pp on unit costs and +17.8 to +21.1\pp on margins.
\end{itemize}

Per-prompt evaluation therefore understates leakage under every mechanism we test: with fresh secrets, the margins and unit costs that one prompt appears to protect are recovered far more often after a history of queries. \Cref{sec:regimes} asks whether a reused secret behaves differently.

\section{Fresh versus Reused Secrets}
\label{sec:regimes}

This section answers Q2. We first compare the two regimes under the same attacker and then explain the contrast.

\subsection{Measured contrast}
\paragraph{Scaling} Under scaling, reuse behaves differently from fresh keys (dashed curves in \cref{fig:linked}). Under \Donep the repeated prompts show identical values, so recovery stays at its $n=1$ level. Under \Done reuse gains +8.9 [$-$2.2, 20.6]\pp in unit cost, consistent with later templates showing additional products whose prices inform the shared key (\cref{fig:keyloc}, bottom). Paired by firm and seed, the fresh-minus-reuse difference in the \Done unit-cost gain is +32.2 [22.8, 41.1]\pp.

\paragraph{Order-preserving substitution} For \Dtwo we apply the joint shared-map attacker to both regimes (\cref{fig:hero} and \cref{tab:d2valid}), because the pair-only attacker sees identical values under reuse and its flat curve does not establish that total leakage stays flat. With a reused map the joint attacker gains +0.6 [$-$4.4, 6.1]\pp in margin hits; with fresh maps it gains +30.0 [18.9, 40.6]\pp and, unlike the pair-only attacker, +17.2 [6.1, 28.3]\pp in unit-cost hits. The fresh-minus-reuse difference is +29.4 [18.3, 40.6]\pp for margins and +17.8 [7.8, 27.8]\pp for unit costs; on the fresh draw the margin difference is +25.6 [12.8, 38.3]\pp. Under a reused map A-exact, which models the exact display, gains +5.6 [$-$3.3, 14.4]\pp in margin hits; its interval does not exclude a reuse gain of up to about 14 points, and its numerical accuracy is only partly validated (\cref{sec:strong}). Even against A-exact's reuse gain, the joint attacker's fresh-map gain leaves a fresh-minus-reuse difference of +24.4 [9.4, 38.3]\pp (fresh draw +22.8 [7.8, 37.2]\pp; \cref{S-sec:robust}).

\begin{table*}[t]
\centering
\footnotesize
\setlength{\tabcolsep}{4pt}
\caption{\Dtwo Attackers: Single-Query Advantage and Linked-Query Gains}
\label{tab:d2valid}
\begin{tabular}{@{}llrrrrr@{}}
\toprule
 & & Single query & \multicolumn{4}{c}{Linked queries, product~0} \\
\cmidrule(lr){3-3}\cmidrule(l){4-7}
Attacker & Map & $\Delta$ margin & $n{=}1$ & $n{=}16$ & Margin gain & Cost gain \\
\midrule
Gaussian pair (default \Aplus), default law & fresh & +1.0 [-2.7, +4.5] & 25.0 & 59.4 & +34.4 [+23.9, +45.0] & +6.1 [-3.3, +15.0] \\
Reference likelihood, default law & fresh & +1.8 [-1.0, +4.4] & 26.1 & 69.4 & +43.3 [+33.3, +53.3] & +7.8 [+1.1, +14.4] \\
Gaussian pair, Dirichlet(0.1) gaps & fresh & -3.0 [-7.3, +1.3] & 17.2 & 29.4 & +12.2 [+1.7, +22.2] & 0.0 [-7.8, +7.8] \\
Reference likelihood, Dirichlet(0.1) gaps & fresh & -0.5 [-2.7, +1.6] & 21.1 & 29.4 & +8.3 [+1.7, +15.6] & +5.6 [-2.8, +13.9] \\
Joint shared-map attacker & fresh & -- & 23.9 & 53.9 & +30.0 [+18.9, +40.6] & +17.2 [+6.1, +28.3] \\
Joint shared-map attacker & reuse & -- & 23.9 & 24.4 & +0.6 [-4.4, +6.1] & -0.6 [-10.6, +8.9] \\
Joint, fresh draw & fresh & -- & 32.2 & 55.0 & +22.8 [+11.1, +35.0] & -2.2 [-12.8, +8.9] \\
Joint, fresh draw & reuse & -- & 32.2 & 29.4 & -2.8 [-8.9, +2.8] & -7.8 [-16.7, +0.6] \\
A-exact (particle, display-exact model) & reuse & -- & 24.4 & 30.0 & +5.6 [-3.3, +14.4] & -- \\
A-exact, fresh draw & reuse & -- & 33.3 & 33.3 & +0.0 [-9.4, +9.4] & -- \\
\bottomrule
\end{tabular}
\par\vspace{2pt}\parbox{\textwidth}{\footnotesize\raggedright Single query: margin advantage over \Azero on the 2{,}100 product rows of the test split (single-query results of the joint attacker and A-exact, on product-0 rows, are in \cref{sec:single}). Linked: product-0 margin hit rate (\%) at $n=1$ and 16 and gains (pp); 60 test firms $\times$ 3 seeds, original firms unless noted. Firm-clustered 95\% bootstrap CIs. A-exact: sampler seed 4646; unit-cost results not shown.}
\end{table*}

\subsection{Why fresh secrets accumulate and reuse does not}
We explain the contrast in two steps. A reused secret caps what the history discloses (\cref{thm:reuse}); fresh secrets are not capped, and whether they leak depends on the mechanism (\cref{thm:cases,thm:rate} and \cref{prop:rank}). Write $X=(X_f)_{f\in\Fset}$ for the firm's values, $F_j$ for the fields shown in query $j$, $F_{\le n}=\bigcup_{j\le n}F_j$, and $T(K,X_F)$ for the full-precision disguised values of fields $F$ under secret $K$. For a target $g(X)$ and loss $\ell\ge0$, let $r(V)=\inf_\delta\E\,\ell(g(X),\delta(V))$ be the Bayes risk of an attacker who sees $V$. Proofs are in \cref{sec:proofs}.

\begin{theorem}[Reuse ceiling]\label{thm:reuse}
Suppose the field sets $F_1,\dots,F_n$ are fixed in advance, the disguise is field-local ($T(K,X_F)$ is the restriction of $T(K,X_\Fset)$ to $F$ for every $F\subseteq\Fset$), each shown query is $Y_j=\rho_j(T(K,X_{F_j}))$ for a measurable display map $\rho_j$, and one secret $K$, independent of $X$, serves all $n$ queries. Then for every target and loss, $r(Y_1,\dots,Y_n)\ge r\big(T(K,X_{F_{\le n}})\big)$.
\end{theorem}

The views in \cref{thm:reuse} are the numeric transcript; the prose and the choice of fields are outside the theorem. Reuse thus bounds $n$ queries with fixed field sets by one full-precision view of all fields shown. That view can carry more than the first prompt, and it retains its invariants: it contains every \Done margin exactly, which the attacker recovered within $\pm5$\pp despite cent rounding (\cref{sec:single}). Fresh secrets need not obey this bound; we consider two cases with independent, identically distributed keys and exact observations.

\begin{theorem}[Two cases for fresh keys]\label{thm:cases}
\textbf{(a)} Let a compact group $G$ act jointly measurably on the value space $\Xset$, let query $j$ show $Y_j=G_j\cdot X$ with $G_1,G_2,\dots$ independent, distributed according to the Haar probability of $G$, and independent of $X$, and let $J$ be a measurable maximal invariant (every measurable $G$-invariant function is $\sigma(J)$-measurable). Then $(Y_1,\dots,Y_n)\indep X\mid J(X)$ for every $n$.
\par\noindent\textbf{(b)} Let $\theta\in\R^d$ be the log-values, $S\subseteq\R^d$ a linear subspace, and $Y_j=\theta+K_j$ with $K_1,K_2,\dots$ independent, distributed according to any Borel probability $\mu$ on $S$, and independent of $\theta$. Then $\theta\mapsto\mathcal{L}(\theta+K_1)$ is injective, and for any Borel prior $\Pi$ the posterior given $Y_1,\dots,Y_n$ converges weakly to $\delta_\theta$ almost surely, for $\Pi$-almost every $\theta$.
\end{theorem}

Case (a) contains modular additive masking, whose maximal invariant is constant, which stays perfectly secret at every $n$ but does not preserve real-valued arithmetic. Case (b) covers the full-precision versions of \Done ($S$ spanned by the all-ones vector of log money fields) and \Donep (one coordinate per field): for fixed positive money fields observed exactly in every query, such as product~0 in our sequences, the ideal posterior under prior $\Pi$ concentrates at the true values for $\Pi$-almost every firm. Neither case covers \Dtwo or \Dthreeo, whose growth in \cref{fig:linked} is an empirical result; \cref{prop:rank} below identifies the property of \Dtwo that matters. For the key law used by \Done, the rate is explicit.

\begin{theorem}[Localization of log-uniform scaling keys]\label{thm:rate}
Let $0<a<b<\infty$, $n\ge1$, and $y_j=k_jx$ with $\log k_1,\dots,\log k_n$ independent and uniform on $[\log a,\log b]$, $L=\log(b/a)$, and exact observations of a fixed $x>0$. The values of $\log x$ consistent with $y_1,\dots,y_n$ form an interval of width $W_n=L-R_n$, where $R_n$ is the range of $\log k_1,\dots,\log k_n$, and $\E[W_n]=2L/(n+1)$.
\end{theorem}

\begin{figure}[t]
  \centering
  \includegraphics[width=0.92\columnwidth]{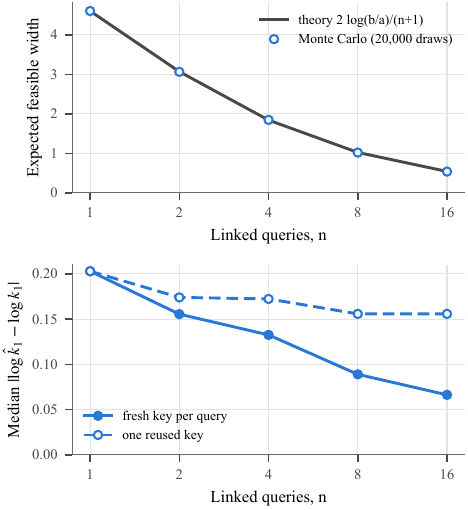}
  \Description{Top: expected feasible log-interval width against the number of queries, theory curve 2 log(b/a)/(n+1) and Monte Carlo points coinciding, falling from 4.6 to 0.5. Bottom: the attacker's median absolute log error of the first key, falling with fresh keys and falling much less with a reused key.}
  \caption{Shared scaling with $k\sim\mathrm{LogUniform}[0.1,10]$. Top: expected width of the feasible log-interval, $2\log(b/a)/(n{+}1)$ (\cref{thm:rate}), against Monte Carlo (20{,}000 draws per $n$). Bottom: median $|\log\hat k_1-\log k_1|$ of the multi-query attacker for the first query's key, fresh vs reused key (60 test firms $\times$ 3 seeds).}
  \label{fig:keyloc}
\end{figure}

With $a=0.1$ and $b=10$ the expected width falls from 4.61 at $n=1$ to 0.54 at $n=16$, matching Monte Carlo (\cref{fig:keyloc}). This is the width of the feasible set under exact observations, not an estimation error; the prior concentrates the posterior within that set (\cref{sec:proofs}, Observation), and the attacker's median key error is 0.203 already at $n=1$.

\paragraph{Which property of \Dtwo leaks} Order preservation alone does not force accumulation.
\begin{proposition}[Order-only maps]\label{prop:rank}
Let $R(X)$ be the rank pattern (order and ties) of all values that the queries pass through the map, let the field sets be fixed in advance, and let query $j$ show $Y_j=g_j(R(X),U_j)$ for a measurable $g_j$ and secrets $U_1,U_2,\dots$ that are independent of each other and of $X$. Then $(Y_1,\dots,Y_n)\indep X\mid R(X)$ for every $n$.
\end{proposition}

A substitution that sorts the distinct values it maps and assigns them independently drawn sorted targets has this form, so any number of linked queries reveals the order and nothing more; this is the ideal notion of order-preserving encoding~\cite{popa2013mope}. Our \Dtwo instead ties each value's target distribution to its position on a public signed-log axis, so each fresh map is an independent noisy reading of that position. Zhang et al.\ sort values into intervals and map them to random targets while preserving order~\cite{zhang2026collaborative}. Which case applies to their construction depends on whether the distribution of its transcript depends on the values beyond their rank pattern; fixing the intervals on a public axis, as our instantiation does, creates such dependence.

\paragraph{How the map law matters} Random order-preserving functions are known to leak approximate values and distances, not only order~\cite{boldyreva2011ope}; the map law sets how much. Varying the law, with the attacker told each law (\cref{S-tab:d2law}), the margin gain from $n=1$ to $16$ stays significant under every law we tried (+12.2 to +51.1\pp). It grows as the map becomes more regular, with more intervals or more even gaps. Under the most regular law (Dirichlet(10) gaps) even a single query gives a margin advantage of +28.0 [23.6, 32.5]\pp, so how much one \Dtwo prompt leaks is a property of the map law rather than of order preservation.

The answer to Q2 is that, for the disguises we study, freshness largely determines what the history discloses: a reused secret caps it at one view, whose invariants can still leak, while fresh secrets give repeated independent views that identify scaled values and locate values mapped on a public axis, unless every view depends on the values through an invariant alone.

\section{Do the Conclusions Depend on the Attacker or the Population?}
\label{sec:bounds}

Every rate above comes from one firm sample, one synthetic generator and concrete attackers. This section answers Q3: whether the conclusions change with the population, the attacker's prior, drifting values, stronger numeric attackers, or an attacker that reads the prose.

\subsection{Replication and population shifts}
\label{sec:robust}
\begin{figure*}[t]
  \centering
  \includegraphics[width=0.68\textwidth]{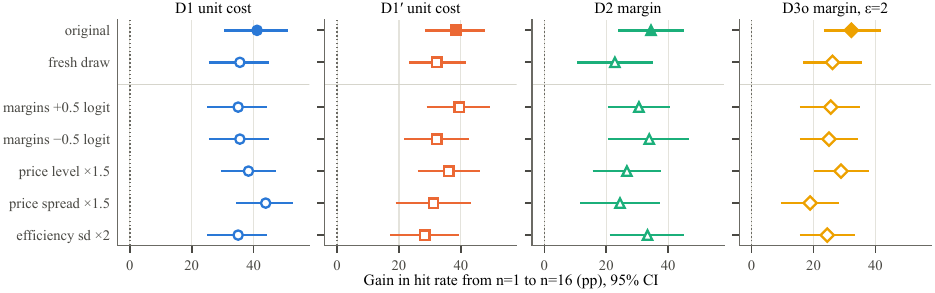}
  \Description{Forest plot in four panels (D1 unit cost, D1-prime unit cost, D2 margin, D3o margin at epsilon 2). Each panel shows the gain in hit rate from 1 to 16 linked queries with its 95 percent interval for the original firms, a fresh draw of firms and five population shifts; every interval lies to the right of zero.}
  \caption{Every fresh-secret gain stays above zero under replication and population shift. Gains for the default attacker \Aplus. Gain in hit rate from $n{=}1$ to $16$ with a fresh secret per query, for the original firms (filled), a fresh draw of firms (new generator seed, same prior table, frozen attacker) and five shifts that each change one generator constant of the fresh draw (open). 60 test firms $\times$ 3 seeds per estimate; bars: firm-clustered 95\% bootstrap CIs; values in \cref{S-tab:robust}.}
  \label{fig:robust}
\end{figure*}

On the fresh draw of firms (\cref{sec:protocol}), the fresh-secret gains replicate (\cref{fig:robust}) except \Dthreeo's margin gain at $\varepsilon=0.5$ (+4.4 [$-$3.3, 12.2]\pp). The conclusions also persist with 20{,}000 instead of 4{,}000 prior draws and under five shifts of the evaluated population with the attacker frozen (\cref{fig:robust}), where every reported linked-query gain (\Done and \Donep unit cost, \Dtwo margin, \Dthreeo margin at $\varepsilon=2$) stays above zero (smallest lower endpoint +9.4\pp). On all 250 test firms (default attacker, $n=1$ and 16), every fresh-secret gain is above zero, including \Dthreeo's margin gain at $\varepsilon=0.5$ (+6.1 [2.8, 9.5]\pp). By sector (41 or 42 firms each), the \Done, \Donep and \Dtwo gains are above zero in all six sectors, whereas \Dthreeo's margin gain is near zero for bakery items at every $\varepsilon$: their prices of a few dollars fall in a domain that extends to 1{,}000, and the noise swamps them (\cref{S-tab:sector}).

\subsection{Prior mismatch and value drift}
\label{sec:mismatch}
\paragraph{Prior from another population model} We rebuild the attacker's prior table from 2{,}000 firms of three alternative models and keep the evaluated firms: lighter-tailed prices with wider location and efficiency spreads and no extreme firms (PM1), randomly shifted class prices and sector margins (PM2), and both (PM3; details in \cref{S-sec:robust}). The single-prompt margin advantages stay at most +4.3\pp. Every fresh-secret history gain stays above zero except \Dthreeo's at $\varepsilon=0.5$, whose interval includes zero under all three models (\cref{S-tab:r049}). Under PM3 the joint attacker's fresh-minus-reuse difference is +31.1 [19.4, 42.2]\pp.

\paragraph{Values that drift} Let each query show a firm's prices and costs multiplied by independent factors with log-standard deviation $\delta$, and let the frozen attacker ignore the drift. The history gains of \Done, \Donep, \Dtwo and \Dthreeo at $\varepsilon\ge1$ stay above zero at every tested $\delta$ up to 5\%, while \Dthreeo at $\varepsilon=0.5$ loses its gain already at $\delta=1\%$ (\cref{S-tab:r049}). Under \Dthreeo a drifted value can fall into another domain, and about 12\% of the $\delta=5\%$ sequences then revert to the prior, so a drift-aware attacker could recover more. \Cref{thm:reuse} assumes unchanged values; reuse under drift is not evaluated.

\subsection{Stronger numeric attackers}
\label{sec:strong}
\paragraph{Reference and joint attackers} The reference likelihood raises the default-law margin gain from +34.4 to +43.3 [33.3, 53.3]\pp (\cref{tab:d2valid}), so at $n=16$ the gain of \cref{fig:linked} is conservative relative to it, although the two curves cross at intermediate $n$. Across fourfold Monte Carlo budgets and smoothings from 0.01 to 0.1, the joint attacker's fresh-minus-reuse margin difference stays between +30.6 and +34.4\pp with every interval above zero, and its reuse gain is at most +5.6 [$-$1.1, 12.2]\pp (\cref{S-tab:numval}).

\paragraph{Numerical accuracy of A-exact} A-exact's two sampler seeds agree within 3.3\pp in every hit rate, and a fourfold budget changes its reuse margin gain by $-1.7$ [$-$5.0, 0.0]\pp (\cref{S-tab:numval}). It is only approximately converged, so we report its margins only (\cref{S-sec:robust}).

\subsection{Do language-model priors help?}
\label{sec:llm}
Beyond the mechanism and a sector prior, a provider can exploit the prose. We test whether reading it with a language model adds substantially to what the numbers leak. A hybrid attacker A4, run with Llama-3.1-8B-Instruct~\cite{grattafiori2024llama3} and with Qwen2.5-32B-Instruct under 4-bit activation-aware weight quantization~\cite{qwen2025qwen25}, reads the masked prose, returns the sector, the product classes, the city's price level and the price effect of the firm's positioning, and shifts the prior table before the frozen \Aplus solver runs. A4-TL keeps the models' cue shifts but replaces their sector and class answers by the truth.

On the development prompts the models read the cues: the Spearman correlation of their city price levels with the regional price parities is 0.63 for Llama and 0.53 for Qwen. On the test split, however, the cue shifts change hit rates by a few points. A4's unit-cost differences to \Aplus under \Done and under \Dthree at $\varepsilon=0.5$ range from $-2.3$ to +2.4\pp, only Llama's +2.4 [0.7, 4.3]\pp has an interval above zero, and A4-TL reaches at most +2.5 [0.8, 4.3]\pp (\cref{S-sec:llm}). The cues carry little in our generator, where the city explains little of a price relative to the spread within a product class: the true city level, fed into the same solver, adds at most +2.3 [0.8, 4.0]\pp. Two further attackers that answer directly from the prompt stay below the prior-only attacker under \Donep and \Dtwo.

Before the test run we had fixed two endpoints, C1a (unit-cost hits under \Done) and C1b (unit-cost or margin hits under \Dthree at $\varepsilon=0.5$), and the criterion of a positive A4$-$\Aplus difference with an interval above zero for both models; only Llama on C1b met it, so the criterion was not met. On these counterfactual firms and with these two open-weight models, the numbers and the query history drive the large effects of \cref{sec:single,sec:linked,sec:regimes}.

Across these tests, the conclusions are not an artifact of a weak attacker or of one sample; they remain conditional on the synthetic population model (\cref{sec:discussion}).

\section{Proofs}
\label{sec:proofs}\label{app:proofs}
This section proves the results of \cref{sec:regimes}.

\paragraph{Notation} $X$ is the firm's secret vector on a standard Borel space with prior $\Pi$; secrets are independent of $X$, and $r(V)$ is the Bayes risk of \cref{sec:regimes}.

\begin{proof}[Proof of \cref{thm:reuse}]
Let $V=T(K,X_{F_{\le n}})$. Because $F_j\subseteq F_{\le n}$ and the same $K$ is used, field locality gives $T(K,X_{F_j})=\mathrm{pr}_{F_j}(V)$, where $\mathrm{pr}_{F_j}$ is the coordinate projection. Hence $(Y_1,\dots,Y_n)=\Phi(V)$ with the measurable map $\Phi=(\rho_1\circ\mathrm{pr}_{F_1},\dots,\rho_n\circ\mathrm{pr}_{F_n})$. For any estimator $\delta$ of $g(X)$ from $(Y_1,\dots,Y_n)$, $\delta\circ\Phi$ is an estimator from $V$ with the same risk, so the infimum over estimators from $V$ is no larger. This is a special case of Blackwell's comparison of experiments~\cite{blackwell1951comparison}.
\end{proof}

\begin{proof}[Proof of \cref{thm:cases}(a)]
A compact group is unimodular, so $\mu$ is right-invariant: $G_1h\sim\mu$ for every $h\in G$. Hence $G_1\cdot(h\cdot x)=(G_1h)\cdot x$ has the same law $P_x$ as $G_1\cdot x$, and the views given $X=x$ have law $P_x^{\otimes n}=P_{h\cdot x}^{\otimes n}$. For a measurable set $B$, $k_B(x)=P^{\otimes n}_x(B)$ is measurable by joint measurability of the action and Fubini's theorem, and $G$-invariant, hence $k_B=\tilde k_B\circ J$ for a measurable $\tilde k_B$. Since the secrets are independent of $X$, $\Prob((Y_1,\dots,Y_n)\in B\mid X)=\tilde k_B(J(X))$ almost surely; being $\sigma(J(X))$-measurable, it also equals $\Prob((Y_1,\dots,Y_n)\in B\mid J(X))$, for every $B$.
\end{proof}

\begin{lemma}\label{lem:noinv}
Let $S\subseteq\R^d$ be a linear subspace and $\Delta\in S\setminus\{0\}$. No Borel probability $\mu$ on $S$ satisfies $\mu(B-\Delta)=\mu(B)$ for every Borel $B\subseteq S$.
\end{lemma}
\begin{proof}
Let $\varphi(t)=\int e^{i\langle t,k\rangle}\mu(dk)$ for $t\in S$. Invariance gives $\varphi(t)=e^{i\langle t,\Delta\rangle}\varphi(t)$. For $t_s=s\Delta/\|\Delta\|^2$ we have $\langle t_s,\Delta\rangle=s$, so $\varphi(t_s)=0$ whenever $0<|s|<2\pi$. But $\varphi$ is continuous by dominated convergence and $\varphi(0)=1$, so $\varphi(t_s)\to1$ as $s\to0$, a contradiction.
\end{proof}

\begin{proof}[Proof of \cref{thm:cases}(b)]
Let $\theta\neq\theta'$. If $\theta-\theta'\notin S$, let $\mathrm{pr}_\perp$ be the orthogonal projection onto $S^\perp$; then $\mathrm{pr}_\perp Y=\mathrm{pr}_\perp\theta$ almost surely, so $P_\theta$ and $P_{\theta'}$ are concentrated on the disjoint sets $\{\mathrm{pr}_\perp y=\mathrm{pr}_\perp\theta\}$ and $\{\mathrm{pr}_\perp y=\mathrm{pr}_\perp\theta'\}$. If $\Delta=\theta-\theta'\in S\setminus\{0\}$, regard $\mu$ as a probability on $\R^d$ concentrated on $S$, so that $P_\theta(C)=\mu(C-\theta)$ for Borel $C\subseteq\R^d$. For Borel $B\subseteq S$ put $C=B+\theta'$: then $P_{\theta'}(C)=\mu(B)$ and $P_\theta(C)=\mu(B-\Delta)$. Equality of the laws would make $\mu$ invariant under translation by $\Delta$, which \cref{lem:noinv} excludes. So $\theta\mapsto P_\theta$ is injective.
For consistency we apply Doob's theorem in the form of~\cite[Theorem~6.9 and Proposition~6.10]{ghosal2017fundamentals}: the parameter and sample spaces $\R^d$ are Polish; $\theta\mapsto P_\theta(A)=\int\ind\{\theta+k\in A\}\mu(dk)$ is measurable by Fubini's theorem; $\theta\mapsto P_\theta$ is injective; and the $Y_j$ are independent with law $P_\theta$ given $\theta$. Hence the posterior is strongly consistent at $\Pi$-almost every $\theta$.
\end{proof}

\begin{proof}[Proof of \cref{thm:rate}]
Write $\theta=\log x$, $u_j=\log y_j=\theta+\kappa_j$ with $\kappa_j=\log k_j$, $\alpha=\log a$ and $\beta=\log b$. The density of $u_j$ given $\theta$ is $L^{-1}\ind\{u_j-\beta\le\theta\le u_j-\alpha\}$, so by independence the likelihood is
\begin{align*}
&\prod_{j=1}^nL^{-1}\ind\{u_j-\beta\le\theta\le u_j-\alpha\}\\
&\qquad=L^{-n}\ind\{\max_ju_j-\beta\le\theta\le\min_ju_j-\alpha\}.
\end{align*}
The consistent values form the interval $I_n=[\max_ju_j-\beta,\ \min_ju_j-\alpha]$, which contains the true $\theta$. Its width is $L-(\max_ju_j-\min_ju_j)=L-R_n$, because the common $\theta$ cancels in $\max_ju_j-\min_ju_j=\max_j\kappa_j-\min_j\kappa_j$. Writing $\kappa_j=\alpha+LV_j$ with $V_j$ independent and uniform on $[0,1]$, $\E\max_jV_j=n/(n+1)$ and $\E\min_jV_j=1/(n+1)$ (Beta$(n,1)$ and Beta$(1,n)$ means), so $\E R_n=L(n-1)/(n+1)$ and $\E W_n=2L/(n+1)$.
\end{proof}

\begin{proof}[Proof of \cref{prop:rank}]
For a measurable set $B$, independence of the secrets from $X$ and of each other gives, by Fubini's theorem, $\Prob((Y_1,\dots,Y_n)\in B\mid X)=\tilde k_B(R(X))$ almost surely, where $\mu_j$ is the law of $U_j$ and
\[\tilde k_B(r)=\textstyle\int\ind_B(g_1(r,u_1),\dots,g_n(r,u_n))\,\mu^{\otimes}(du),\]
with $\mu^{\otimes}=\mu_1\otimes\cdots\otimes\mu_n$, is measurable. Being $\sigma(R(X))$-measurable, it also equals $\Prob((Y_1,\dots,Y_n)\in B\mid R(X))$, for every $B$.
\end{proof}

\paragraph{Observation (posterior support)} Under the assumptions of \cref{thm:rate}, let the prior on $\theta$ have a density $f$ that is positive everywhere. Since the likelihood equals $L^{-n}\ind\{\theta\in I_n\}$ and $I_n$ has positive length almost surely, the posterior density is, almost surely, $f(\theta)\ind\{\theta\in I_n\}/\int_{I_n}f$, which is positive exactly on $I_n$. The posterior support of $\log k_1=u_1-\theta$ is $u_1-I_n$, of the same width, and any other money field shown under the same keys is a deterministic function of $\theta$ given the views.

\section{Discussion}
\label{sec:discussion}

\paragraph{Three checks for disguise designers} First, evaluate a disguise against the history it produces, not one prompt (\cref{sec:single,sec:linked}). Second, choose between fresh and reused secrets deliberately: for field-local mechanisms, a reused secret for unchanged fields bounds history leakage (\cref{thm:reuse}) at the price of what the union view exposes, such as every margin under shared scaling. Third, test whether outputs depend on the values beyond an invariant such as their order: if not, fresh randomness discloses only that invariant, which may itself be sensitive (\cref{thm:cases}a, \cref{prop:rank}), whereas a public axis or an input-dependent domain adds such dependence.

\paragraph{Recommendations for the evaluated mechanisms} For released \preempt, caching one sanitized value per firm and field would make every later query a function of that release, and a domain fixed in advance would remove the input-dependent support (\cref{sec:background}); we have not evaluated either.

\paragraph{Scope and limitations} First, all firms are synthetic: sector margins are anchored to public aggregates, but prices, their dispersion and the cue-to-price link are assumptions, and \cref{sec:bounds} varies the generator and the attacker's prior but evaluates no real firm. Second, \Dtwo is our public-axis instantiation (\cref{tab:d2fid}); Zhang et al.'s topic shift, the utility of any defense and a total privacy budget per firm are not evaluated. Third, our attackers use approximate likelihoods and leave within-interval relations unused under fresh maps, so an optimal attacker may recover more in either regime.

\paragraph{Ethics} The only real data is the public FinQA benchmark, and no personal data or human subjects are involved; we describe the input-dependent domain of released \preempt's public code so that its deployments can be checked.

\section{Conclusion}
\label{sec:conclusion}

Numeric prompt disguises are evaluated one prompt at a time, but a provider sees a firm's history of queries. We have shown that this history matters: disguises that appear to protect margins in one prompt leak them as linked queries with fresh secrets accumulate (\preempt at $\varepsilon\ge0.5$), whereas a reused secret caps the history at one full-precision view of the fields shown, whose invariants can still leak; under the order-preserving substitution, one joint attacker gains 29.4\pp more in margin hits with fresh maps than with one reused map. Disguises for repeatedly queried numbers should therefore be evaluated against the history they produce, with the choice between fresh and reused secrets made deliberately. Future work should test the results on disclosed product-line margins, evaluate the suggested mitigations, and measure Zhang et al.'s pipeline.

\clearpage
\setcounter{section}{0}
\setcounter{table}{0}
\setcounter{figure}{0}
\renewcommand{\thesection}{S\arabic{section}}
\renewcommand{\thesectiondis}{S\arabic{section}.}
\renewcommand{\thetable}{S\arabic{table}}
\renewcommand{\thefigure}{S\arabic{figure}}
\renewcommand{\theHsection}{S\arabic{section}}
\renewcommand{\theHtable}{S\arabic{table}}
\renewcommand{\theHfigure}{S\arabic{figure}}
\renewcommand{\topfraction}{0.95}
\renewcommand{\dbltopfraction}{0.7}
\renewcommand{\textfraction}{0.05}
\renewcommand{\floatpagefraction}{0.85}
\renewcommand{\dblfloatpagefraction}{0.85}
\setcounter{topnumber}{4}
\setcounter{dbltopnumber}{2}
\setcounter{totalnumber}{6}
\twocolumn[{\centering\LARGE Supplemental Material\par\vspace{1.5em}}]

\noindent This supplemental material holds the data, mechanism and attacker details, further single-prompt and robustness results, and the language-model study of the main paper; the code release (\url{https://github.com/Jamli58/fresh-keys-stale-secrets}) includes an evidence index that maps every result to its file. Its sections, tables and figures are numbered S1, S2, \ldots; references to Sections I--XI, Theorems 1--3, Proposition 4, Tables I--V and Figs.~1--6 refer to the main paper.

\section{Data}
\label[appendixsec]{S-sec:data}
\paragraph{Sector margin anchors} \Cref{S-tab:generator} lists the generator's sector margin medians with the other constants. For the five sectors anchored to IRS Statistics of Income~\cite{irs2022soi}, the gross margin is one minus cost of goods sold over business receipts for the listed minor industry (returns of active corporations, tax year 2022). For IT consulting we use the January 2026 Computer Services gross margin (24.26\%)~\cite{damodaran2026margins} as a modeling proxy instead of the SOI ratio for computer systems design (0.69), which we judged too high for a consulting day's margin.

\paragraph{Generator} Product prices are a class median times the firm's location factor $\mathrm{RPP}/100\cdot e^{\eta+\beta}$, with city residual $\eta\sim N(0,0.08^2)$ and brand premium $\beta\sim N(0,0.15^2)$, times Student-$t$ noise with 3 degrees of freedom on the log scale, scaled to the class's log-price spread. Margins are the logistic transform of the sector's logit median plus the firm's efficiency $e$ ($N(0,0.3^2)$, or $\pm$1.2--2.0 for firms drawn as extreme with probability 0.2; 22.3\% of the 300 firms) plus logit noise with the sector's spread, clipped to $[0.01,0.97]$. Each firm sells 3 to 8 products, and monthly volumes carry a seasonal amplitude of 0.10. Cities are 60 metropolitan areas taken at evenly spaced ranks of the 2024 Regional Price Parities~\cite{bea2024rpp}. Positioning is premium if $s>0.8$, discount if $s<-0.8$ and standard otherwise, for the score $s=\beta/0.15+0.5\,e/0.3+N(0,0.7^2)$; in the 300 firms the mean log price residual is $-0.16$ for discount, $+0.04$ for standard and $+0.02$ for premium firms. \Cref{S-tab:generator} lists the per-sector constants.

\begin{table*}[!t]
\centering
\footnotesize
\caption{Generator Constants per Sector}
\label{S-tab:generator}
\resizebox{\textwidth}{!}{%
\begin{tabular}{@{}lrrrrrl@{}}
\toprule
Sector & Margin median & Margin logit sd & Volume median & Volume log-sd & Overhead median & Product classes \\
\midrule
Bakery/caf\'e & 0.580 & 0.35 & 900 & 0.6 & 18{,}000 & bread 4.5 (0.35), pastry 3.8 (0.30), coffee 4.2 (0.25), cake 32 (0.40) \\
Machine shop & 0.360 & 0.35 & 60 & 0.8 & 45{,}000 & machined part 85 (0.8), fixture 420 (0.6), repair job 260 (0.5) \\
Apparel e-commerce & 0.450 & 0.35 & 300 & 0.9 & 25{,}000 & T-shirt 24 (0.35), jacket 120 (0.45), accessory 18 (0.5) \\
IT consulting & 0.240 & 0.40 & 25 & 0.7 & 60{,}000 & consulting day 950 (0.35), support contract 1{,}500 (0.5), licence 400 (0.6) \\
Construction sub. & 0.340 & 0.35 & 40 & 0.8 & 70{,}000 & m$^2$ install 55 (0.4), site day 1{,}400 (0.35), materials lot 3{,}200 (0.6) \\
Specialty retail & 0.385 & 0.35 & 150 & 0.8 & 20{,}000 & small item 15 (0.5), mid item 65 (0.45), premium item 240 (0.5) \\
\bottomrule
\end{tabular}}
\par\vspace{2pt}\parbox{\textwidth}{\footnotesize\raggedright Margin medians: rounded SOI gross margins of food services and drinking places; hardware, spring, wire and machine shops; clothing and accessories retailers; specialty trade contractors; sporting goods, hobby, book and music stores; IT consulting: computer-services industry table. All other constants are documented assumptions. Product classes: median price (log-price spread).}
\end{table*}

\paragraph{Templates} Besides product~0's price and cost, which every template shows, quote pricing adds product~0's volume, the overhead and an order quantity; margin review the prices and costs of up to two more products; break-even the overhead; the sales summary three monthly volumes and a report year; supplier comparison a supplier quote and a minimum order.

\paragraph{Example prompt} The margin-review prompt of the first test firm (apparel e-commerce), plain and disguised (defense seed 0); only the numbers change across disguises:
\begin{quote}\small
\emph{Plain:} Kraesmelpox Works is based in Peoria, IL. We compete on low prices and high volume. We are Kraesmelpox Works, a small Apparel E-commerce business. Here are our main products. Senvas jacket: price 104.03, unit cost 90.01. Lathkael jacket: price 173.46, unit cost 149.54. Lertas jacket: price 82.92, unit cost 74.59. Which product has the weakest margin, and what would you change first?\\
\emph{\Done:} \dots\ price 195.47, unit cost 169.13 \dots\ price 325.93, unit cost 280.98 \dots\ price 155.81, unit cost 140.15 \dots\\
\emph{\Donep:} \dots\ price 195.47, unit cost 31.18 \dots\ price 20.95, unit cost 16.14 \dots\ price 350.91, unit cost 499.10 \dots\\
\emph{\Dtwo:} \dots\ price 4,509.19, unit cost 4,416.86 \dots\ price 4,851.98, unit cost 4,749.76 \dots\ price 4,365.44, unit cost 4,300.04 \dots\\
\emph{\Dthreeo, $\varepsilon=1$:} \dots\ price 298.00, unit cost 81.00 \dots\ price 100.00, unit cost 100.00 \dots\ price 101.00, unit cost 111.00 \dots
\end{quote}

\section{Mechanism Details}
\label[appendixsec]{S-sec:mech}
\paragraph{Released money sanitizer (\Dthreeo)} \Dthreeo implements \texttt{encypt\_money} and \texttt{M\_epsilon} of \preempt's released code~\cite{preemptcode} (\cref{sec:background}). The domains are $[1,2]$ if $v=1$, $[2,1000]$ if $v<100$, $[100,10^4]$ if $v<1000$ and $[1000,10^4]$ otherwise, so one bin is about 10 dollars for values from 2 to 99; each money value receives the full $\varepsilon$, counts and years are unchanged, and in the linked-query experiments every release draws its own noise. Within a domain, $\Pr[s\mid v]=e^{-|v'-s|\varepsilon/2}/Z(v)$ with $Z(v)=\sum_{t}e^{-|v'-t|\varepsilon/2}$; by the triangle inequality the numerator and $Z$ change by at most a factor $e^{|v'_1-v'_2|\varepsilon/2}$ between $v_1$ and $v_2$, which gives the metric-DP inequality~\cite{chatzikokolakis2013broadening,andres2013geo}. Across domains some outputs are possible for one input and impossible for the other, so the ratio is unbounded. Supports of different domains can overlap (1090 lies on the output grids of both $[100,10^4]$ and $[1000,10^4]$), so an output narrows down the domain without always identifying it; our likelihood sums the bin probabilities of the shown output within the value's domain.

\paragraph{Our variant (\Dthree)} \Dthree follows the paper's description: each of the $N_{\mathrm{v}}$ value tokens of a prompt receives $\varepsilon'=\varepsilon/N_{\mathrm{v}}$ and is replaced by a draw from a truncated-Laplace density with scale $2/\varepsilon'$ on $[0,50{,}000]$ for money, $[0,20{,}000]$ for counts and $[0,100]$ for percentages, with distance in displayed units. Equal nominal $\varepsilon$ is therefore not an equal budget for \Dthreeo and \Dthree.

\paragraph{\Dtwo text-level rules} The year and special-number rules matter on real financial text: 87\% of the 1{,}108 FinQA arithmetic test items contain a year-range integer and 38\% an integer from 28 to 31.

\section{Attacker Details}
\label[appendixsec]{S-sec:attackers}
\paragraph{Classical attacker} The \Aplus prior shape (normal or Student-$t$ with 3 degrees of freedom) and width factor (0.75, 1, 1.5) were chosen on the development split by the average of margin and unit-cost hit rates and then frozen. The likelihood weights the prior draws: \Donep up to whole-cent rounding (the key interval consistent with a half-cent error), \Dthreeo by the released mass function, \Dthree by its truncated-Laplace density and \Dtwo by the bivariate Gaussian of \cref{sec:framework}. For hit-optimal decisions the estimate maximizes the posterior mass of the hit interval, a sliding window over the weighted draws. Under \Done the multi-query attacker uses a grid of 1{,}201 points over $\log k_1$: with $r_j$ the ratio of product~0's shown price in query $j$ to that in query 1, it keeps the keys for which every $k_1r_j$ lies in $[0.1,10]$, to within half a cent, and uses each distinct product's price prior once.

\paragraph{A-exact} \emph{Coordinates.} On the public axis $t=\log(1+x)$, the map $h_\Lambda$ with knot vector $\Lambda$ (log targets at the interval endpoints, drawn from the law $p(\Lambda)$ of \cref{app:mech}) is affine in log space on each interval: for $t$ in interval $j$, of width $w$, $u=\log h_\Lambda(x)=\Lambda_j+(t-t_j)(\Lambda_{j+1}-\Lambda_j)/w$. A-exact works in the latent outputs $u_i$ of the shown mapped fields, so $\partial t_i/\partial u_i=w/(\Lambda_{j+1}-\Lambda_j)$ and, with $z_i=\log x_i$, $\partial z_i/\partial u_i=\frac{1+x_i}{x_i}\cdot\frac{w}{\Lambda_{j+1}-\Lambda_j}$. Because $h_\Lambda$ is strictly increasing, each $u_i$ inverts to a unique $x_i(\Lambda,u)$.
\emph{Display.} \Dtwo rounds $e^{u_i}$ to $d$ significant digits, the smallest $d\ge6$ that keeps distinct inputs of the prompt distinct, so field $i$ is consistent with its shown value $y_i$ exactly when $u_i$ lies in the log rounding cell $C_i$ of $y_i$. All evaluated displays used $d=6$; any $(\Lambda,u)$ with all $u_i\in C_i$ then selects six digits itself, so the precision rule adds no factor beyond the cell indicators, and fields with identical shown values share one latent value.
\emph{Target.} With $\mathcal P$ the products and $\mathcal O$ the other mapped fields,
\begin{align*}
\pi(\Lambda,u)\propto{}&p(\Lambda)\prod_{k\in\mathcal P}g_k(z_{p_k},z_{c_k})\Big|\tfrac{\partial z_{p_k}}{\partial u_{p_k}}\tfrac{\partial z_{c_k}}{\partial u_{c_k}}\Big|\\
&\times\prod_{i\in\mathcal O}\tfrac{1}{t_{\max}}\Big|\tfrac{\partial t_i}{\partial u_i}\Big|\prod_i\ind\{u_i\in C_i\},
\end{align*}
where the factor $\big|\tfrac{\partial z_{p_k}}{\partial u_{p_k}}\tfrac{\partial z_{c_k}}{\partial u_{c_k}}\big|$ is the Jacobian $J_k$ of the main text, $g_k$ is product $k$'s oracle class prior written in $(\log p,\log c)$: Student-$t$ in $(\log p,\operatorname{logit}m)$ as for \Aplus, times $|\partial(\log p,\operatorname{logit}m)/\partial(\log p,\log c)|=1/m$; and each other field has the proper uniform prior on $t\in(0,t_{\max}]$, $t_{\max}=\log(1+10^{12})$. The marginal of $\pi$ in $x(\Lambda,u)$ is the posterior of the values under the exact display model.
\emph{Unmapped fields.} A shown integer in 28--31 is read as its true value and one in $[1980,2040]$ as a shifted year, which A-exact does not use; every other shown value is mapped. A mapped value is misread as a year only if its six-digit display is such an integer, which happened to 11 of 145{,}440 shown values in our evaluation, none in a reused-map sequence.
\emph{Sampling.} The first proposal draws $\Lambda\sim p(\Lambda)$ and each $u_i$ uniformly in $C_i$, so the importance weight is the product of the prior and Jacobian factors above, up to a constant. Under a reused map each particle keeps its map: values seen before must stay in their cells, new values are inverted under the particle's map, and after each query the particles are resampled and moved by Metropolis--Hastings steps on single knots and in-cell positions, with an exact Gibbs refresh of knots that are not adjacent to any value.

\section{Single-Prompt Details}
\label[appendixsec]{S-sec:single}
\paragraph{Typical and atypical firms} The small margin advantages of \Donep, \Dtwo and \Dthreeo (\cref{tab:single}) average over mostly typical firms. On the 498 rows of atypical firms, whose median margin lies outside their sector's P10--P90 band, \Azero hits no margin and \Aplus hits 1.0\% under \Donep, 3.2\% under \Dtwo and 2.0 to 8.2\% under \Dthreeo (\cref{S-tab:strata}). Margin errors still shrink somewhat (mean absolute error 13.3\pp for \Azero versus 11.4\pp under \Dtwo), and unit costs leak slightly under \Donep (+3.4 [1.0, 5.7]\pp).

\begin{table}[!tb]
\centering
\footnotesize
\setlength{\tabcolsep}{3pt}
\caption{Single-Query Recovery by Firm Type}
\label{S-tab:strata}
\begin{tabular}{@{}lrrrr@{}}
\toprule
\multicolumn{5}{@{}l}{(a) Typical firms (1{,}602 rows)} \\
Mechanism & margin & $\Delta$ margin & cost & $\Delta$ cost \\
\midrule
\Done & 100.0 & +63.1 [+57.7, +68.4] & 28.2 & +8.2 [+1.9, +14.4] \\
\Donep & 36.9 & 0.0 [-2.3, +2.3] & 22.3 & +2.2 [-0.7, +5.2] \\
\Dtwo (ours) & 37.1 & +0.2 [-4.4, +4.9] & 18.2 & -1.9 [-5.3, +1.4] \\
\Dthreeo, $\varepsilon$=0.5 & 35.5 & -1.4 [-4.9, +2.1] & 29.6 & +9.6 [+4.9, +14.3] \\
\Dthreeo, $\varepsilon$=1.0 & 38.1 & +1.2 [-2.8, +5.0] & 36.3 & +16.2 [+10.9, +21.6] \\
\Dthreeo, $\varepsilon$=2.0 & 40.1 & +3.2 [-1.1, +7.4] & 42.8 & +22.7 [+17.0, +28.4] \\
\midrule
\multicolumn{5}{@{}l}{(b) Atypical firms (498 rows)} \\
Mechanism & margin & $\Delta$ margin & cost & $\Delta$ cost \\
\midrule
\Done & 100.0 & +100.0 [+100.0, +100.0] & 29.9 & +26.9 [+15.9, +38.5] \\
\Donep & 1.0 & +1.0 [+0.2, +2.1] & 10.0 & +7.0 [+3.8, +10.0] \\
\Dtwo (ours) & 3.2 & +3.2 [+1.5, +5.2] & 8.0 & +5.0 [+0.8, +9.3] \\
\Dthreeo, $\varepsilon$=0.5 & 2.0 & +2.0 [0.0, +4.7] & 14.9 & +11.8 [+6.6, +17.9] \\
\Dthreeo, $\varepsilon$=1.0 & 2.0 & +2.0 [+0.2, +4.7] & 16.1 & +13.1 [+7.5, +19.2] \\
\Dthreeo, $\varepsilon$=2.0 & 8.2 & +8.2 [+4.3, +12.8] & 23.7 & +20.7 [+13.9, +28.0] \\
\bottomrule
\end{tabular}
\par\vspace{2pt}\parbox{\columnwidth}{\footnotesize\raggedright Test split. Atypical: median product margin outside the sector's P10--P90 band of the prior table. Hit rate of \Aplus (\%) and advantage over \Azero (pp, 95\% CI). \Azero hits 36.9\% / 0.0\% of margins and 20.0\% / 3.0\% of unit costs (typical / atypical).}
\end{table}

\paragraph{Released \preempt leaks through its support} In the control $\varepsilon\to0$ ($\varepsilon=10^{-6}$), each release is a draw from the output grid of its value's domain that is uniform to within 0.003\% per bin, so it depends on the value almost only through the domain. The advantage is then +4.8 [0.9, 8.6]\pp for unit costs and $-2.0$ [$-$4.6, 0.4]\pp for margins, below the advantages at $\varepsilon\ge0.5$, where the noise within a domain adds to it.

\paragraph{Stronger attackers on \Dtwo} On the 300 product-0 rows of the margin-review prompts, which show three products, the joint attacker gains +5.0 [$-$1.3, 11.3]\pp in margin hits and +7.3 [0.7, 14.0]\pp in unit-cost hits over \Azero, and A-exact gains +8.7 [0.7, 17.0]\pp in margin hits over the pair-only attacker \Aplus; all 500 test prompts give 1{,}500 product-0 rows.

\paragraph{Decision rule and population} When \Aplus and a prior-only control on the same prior choose the estimate that maximizes the posterior probability of a hit, the margin advantages stay small: +0.4 [$-$0.6, 1.4]\pp under \Donep, +1.4 [$-$1.5, 4.1]\pp under \Dtwo, and 0.0, +1.8 and +4.8 [1.3, 8.2]\pp under \Dthreeo. For \Aplus the two decision rules differ by at most 1.0\pp in margin hit rate and 1.8\pp in cost hit rate, none significantly, and the control's median agrees with \Azero (+0.1 [$-$1.2, 1.4]\pp margin hits). On a fresh draw of firms the margin advantages are 0.0 [$-$1.5, 1.5]\pp (\Donep), $-0.6$ [$-$4.4, 3.1]\pp (\Dtwo) and $-0.9$ [$-$4.0, 2.3]\pp (\Dthreeo, $\varepsilon=1$), and the \Dthreeo unit-cost advantage at $\varepsilon=1$ is +13.4 [9.5, 17.1]\pp. They depend on how well the prior fits: when the evaluated firms' margins are 0.5 logit lower than the prior table assumes, \Azero's margin hit rate falls from 30.4\% to 17.1\%, and the shown numbers add +9.7 [6.2, 13.3]\pp under \Dtwo and +6.1 [3.4, 9.0]\pp under \Dthreeo. Under the other shifts of the fresh draw (margins +0.5 logit, price level or spread $\times1.5$, firm efficiency spread $\times2$; \Azero at 15.7\% for higher margins and 21.7\% for wider efficiency), the \Dtwo and \Dthreeo margin advantages lie between $-1.5$ and +3.0\pp with intervals that include zero, and under all five shifts the \Dthreeo unit-cost advantage stays between +12.6 and +19.9\pp.

\begin{figure}[t]
  \centering
  \includegraphics[width=\columnwidth]{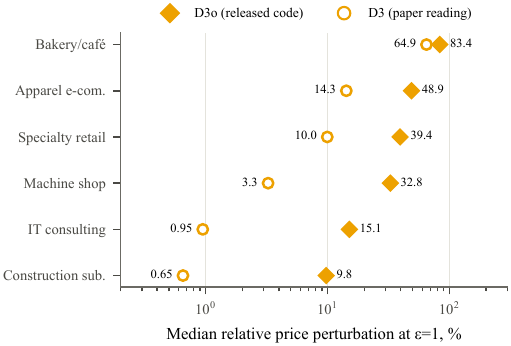}
  \Description{Dot plot, log scale, of the median relative perturbation of a shown price at epsilon 1 for each of six sectors, for D3o and D3; perturbation is largest for bakery items and smallest for construction subcontracting.}
  \caption{Median relative perturbation of the shown price at $\varepsilon{=}1$ by sector (log scale). \Dthreeo: released money sanitizer, full $\varepsilon$ per money value; all 1{,}624 generated products $\times$ 5 draws (8{,}120 observations). \Dthree: our variant of the paper's description, $\varepsilon/N_{\mathrm{v}}$ per value token; products of the evaluation prompts $\times$ 3 seeds (2{,}520 observations). The two populations and budget allocations differ.}
  \label{S-fig:perturbation}
\end{figure}

\paragraph{Released code versus our variant} \Dthreeo and \Dthree differ in domain, binning and budget allocation. In the descriptive summaries of \cref{S-fig:perturbation}, which use different product populations, the released version's median price perturbation is larger in every sector. Both perturb prices relatively less in sectors with large amounts: at $\varepsilon=1$ the median relative perturbation of a shown price is 83.4\% for bakery items but 9.8\% for construction subcontracting under \Dthreeo, and 64.9\% versus 0.65\% under \Dthree, so the same $\varepsilon$ spends very different relative noise across sectors.

\section{Robustness of the Linked-Query Results}
\label[appendixsec]{S-sec:robust}
\paragraph{Replication on a fresh draw} We repeat the $n=1\to16$ comparison on the first 60 test firms of a fresh draw of 300 firms (new generator seed, same prior table, frozen attacker; prior-only baselines 25.0\% for margins and 23.3\% for costs). The fresh-secret gains replicate: \Done unit cost +35.6 [25.6, 45.0]\pp, \Donep unit cost +32.2 [23.3, 41.7]\pp and margin +15.6 [6.1, 25.6]\pp, \Dtwo margin +22.8 [10.6, 35.0]\pp, and \Dthreeo margin +16.7 [8.3, 25.6]\pp at $\varepsilon=1$ and +26.1 [16.7, 35.6]\pp at $\varepsilon=2$; only the \Dthreeo margin gain at $\varepsilon=0.5$ does not replicate significantly (+4.4 [$-$3.3, 12.2]\pp; second row of \cref{S-tab:robust}). The \Dthreeo unit-cost gains are +8.3 [0.6, 16.7], +23.9 [15.0, 32.8] and +20.6 [11.1, 30.6]\pp at $\varepsilon=0.5,1,2$, and \Dtwo shows no unit-cost gain ($-1.1$ [$-$10.6, 8.3]\pp). Under reuse the pair-only attackers stay at their $n=1$ level, since the repeated views are identical, and \Done gains a non-significant +9.4 [$-$0.6, 19.4]\pp.

\paragraph{Prior draws and population shifts} With 20{,}000 instead of 4{,}000 prior draws on the original firms the conclusions persist, for example \Donep margin +18.9 [8.9, 28.3]\pp and \Dtwo margin +35.6 [25.0, 46.1]\pp; for \Dthreeo the margin gains are +7.8 [2.2, 13.9], +17.2 [7.2, 27.2] and +26.1 [16.1, 36.7]\pp at $\varepsilon=0.5,1,2$ (+32.2\pp with 4{,}000 draws at $\varepsilon=2$) and the unit-cost gains +13.3 to +20.6\pp, all intervals above zero. With the attacker frozen and the evaluated population shifted (margins $\pm0.5$ logit, price level or spread $\times1.5$, firm efficiency spread $\times2$), every linked-query gain stays above zero, with the smallest lower endpoint at +9.4\pp (\cref{S-tab:robust}).

\begin{table*}[!t]
\centering
\scriptsize
\setlength{\tabcolsep}{3pt}
\caption{Fresh-Secret Gains under Replication, Population Shift, Prior Mismatch and Value Drift}
\label{S-tab:robust}\label{S-tab:r049}
\begin{tabular}{@{}lrrrrrr@{}}
\toprule
 & \Done cost & \Donep cost & \Dtwo margin & \multicolumn{3}{c}{\Dthreeo margin} \\
\cmidrule(l){5-7}
Setting & & & & $\varepsilon$=0.5 & $\varepsilon$=1 & $\varepsilon$=2 \\
\midrule
original firms & +41.1 [+30.6, +51.1] & +38.3 [+28.3, +47.8] & +34.4 [+23.9, +45.0] & +8.3 [+2.2, +15.0] & +16.7 [+8.3, +25.0] & +32.2 [+23.3, +41.7] \\
fresh draw & +35.6 [+25.6, +45.0] & +32.2 [+23.3, +41.7] & +22.8 [+10.6, +35.0] & +4.4 [-3.3, +12.2] & +16.7 [+8.3, +25.6] & +26.1 [+16.7, +35.6] \\
\midrule
\multicolumn{7}{@{}l}{\emph{Fresh draw with one generator constant changed, attacker frozen}} \\
margins $+0.5$ logit & +35.0 [+25.0, +44.4] & +39.4 [+28.9, +49.4] & +30.6 [+20.6, +40.6] & -- & -- & +25.6 [+15.6, +35.0] \\
margins $-0.5$ logit & +35.6 [+25.6, +45.0] & +32.2 [+21.7, +42.8] & +33.9 [+20.6, +46.7] & -- & -- & +25.0 [+15.6, +34.4] \\
price level $\times1.5$ & +38.3 [+29.4, +47.2] & +36.1 [+26.1, +46.1] & +26.7 [+15.6, +37.8] & -- & -- & +28.9 [+20.0, +37.8] \\
price spread $\times1.5$ & +43.9 [+34.4, +52.8] & +31.1 [+18.9, +43.3] & +24.4 [+11.7, +37.2] & -- & -- & +18.9 [+9.4, +28.3] \\
firm efficiency sd $\times2$ & +35.0 [+25.0, +44.4] & +28.3 [+17.2, +39.4] & +33.3 [+21.1, +45.0] & -- & -- & +24.4 [+15.6, +33.3] \\
\midrule
\multicolumn{7}{@{}l}{\emph{Original firms, attacker prior from another population model}} \\
prior PM1 (tails, spreads) & +41.1 [+31.7, +50.6] & +36.7 [+27.2, +46.1] & +40.0 [+29.4, +50.6] & +6.1 [-0.6, +13.3] & +13.3 [+5.6, +21.7] & +26.1 [+16.1, +36.1] \\
prior PM2 (locations) & +41.7 [+32.2, +50.6] & +38.3 [+28.3, +48.3] & +38.3 [+26.1, +50.0] & +2.8 [-2.2, +7.8] & +13.9 [+3.3, +24.4] & +20.0 [+10.6, +30.0] \\
prior PM3 (both) & +46.7 [+37.2, +55.6] & +38.9 [+29.4, +48.3] & +38.9 [+26.7, +50.6] & -0.6 [-5.6, +3.9] & +12.8 [+5.0, +21.1] & +22.2 [+14.4, +30.6] \\
\midrule
\multicolumn{7}{@{}l}{\emph{Original firms, prices and costs drifting between queries}} \\
drift $\delta=1\%$ & +42.8 [+32.2, +52.8] & +40.6 [+30.6, +50.6] & +35.6 [+25.0, +46.1] & +3.9 [-2.2, +10.0] & +13.9 [+6.7, +21.7] & +27.2 [+17.8, +36.7] \\
drift $\delta=2\%$ & +35.0 [+24.4, +45.6] & +39.4 [+28.3, +50.0] & +35.0 [+24.4, +45.6] & +2.2 [-3.3, +8.3] & +13.3 [+6.1, +21.1] & +28.3 [+19.4, +37.2] \\
drift $\delta=5\%$ & +28.3 [+19.4, +37.2] & +36.1 [+24.4, +47.2] & +33.3 [+22.2, +43.9] & +0.6 [-5.0, +6.7] & +7.2 [+1.7, +13.3] & +21.7 [+11.7, +31.7] \\
\bottomrule
\end{tabular}
\par\vspace{2pt}\parbox{\textwidth}{\footnotesize\raggedright Gain in hit rate (pp) from $n{=}1$ to $16$ with fresh secrets, default attacker \Aplus, 60 test firms $\times$ 3 seeds per cell, firm-clustered 95\% bootstrap CIs. The \Done, \Donep, \Dtwo and \Dthreeo $\varepsilon$=2 columns of the first seven rows are the values of \cref{fig:robust}. Fresh draw: new generator seed, same prior table, frozen attacker. PM1--PM3: the attacker's prior table comes from another population model. Drift: query $j$ shows prices and costs multiplied by independent factors $e^{N(0,\delta^2)}$, which the attacker ignores. --: not run.}
\end{table*}

\paragraph{Sectors} On all 250 test firms (\cref{S-tab:sector}), the \Done, \Donep and \Dtwo gains have intervals above zero in every sector, whereas \Dthreeo's margin gains are near zero for bakery items at every $\varepsilon$ and vary widely across the other sectors. On the 60-firm fresh draw, the \Dtwo and \Dthreeo margin gains for bakery and apparel lie between $-6.7$ and +10.0\pp, with intervals that include zero.

\begin{table*}[!t]
\centering
\scriptsize
\setlength{\tabcolsep}{3pt}
\caption{Linked-Query Gains by Sector on All Test Firms}
\label{S-tab:sector}
\begin{tabular}{@{}lrrrrrr@{}}
\toprule
 & \Done cost & \Donep cost & \Dtwo margin & \multicolumn{3}{c}{\Dthreeo margin} \\
\cmidrule(l){5-7}
Sector & & & & $\varepsilon$=0.5 & $\varepsilon$=1 & $\varepsilon$=2 \\
\midrule
Bakery/caf\'e & +30.1 [+17.1, +42.3] & +36.6 [+26.0, +47.2] & +23.6 [+8.9, +38.2] & +0.0 [-2.4, +2.4] & +0.0 [-2.4, +2.4] & +3.3 [-1.6, +8.1] \\
Machine shop & +37.4 [+23.6, +49.6] & +47.2 [+37.4, +56.1] & +35.0 [+22.0, +48.0] & +3.3 [-4.9, +12.2] & +15.4 [+7.3, +24.4] & +17.9 [+5.7, +30.1] \\
Apparel e-com. & +42.1 [+30.2, +54.0] & +38.9 [+25.4, +51.6] & +35.7 [+22.2, +49.2] & +4.0 [-3.2, +11.9] & +10.3 [+1.6, +19.8] & +12.7 [+0.8, +24.6] \\
IT consulting & +18.3 [+4.8, +31.7] & +50.8 [+39.7, +61.1] & +23.0 [+11.1, +35.7] & +0.0 [-9.5, +9.5] & +23.0 [+10.3, +35.7] & +36.5 [+23.8, +48.4] \\
Construction sub. & +46.0 [+30.2, +60.3] & +44.4 [+34.1, +54.8] & +45.2 [+31.0, +59.5] & +18.3 [+8.7, +28.6] & +26.2 [+15.1, +37.3] & +38.1 [+27.8, +49.2] \\
Specialty retail & +27.8 [+12.7, +42.1] & +40.5 [+27.0, +54.0] & +27.0 [+13.5, +40.5] & +11.1 [+4.0, +19.0] & +9.5 [-1.6, +21.4] & +27.8 [+15.1, +41.3] \\
\bottomrule
\end{tabular}
\par\vspace{2pt}\parbox{\textwidth}{\footnotesize\raggedright Gain in hit rate (pp) from $n{=}1$ to $n{=}16$ with fresh secrets, default attacker \Aplus, all 250 test firms (41 or 42 per sector) $\times$ 3 seeds; firm-clustered 95\% bootstrap CIs.}
\end{table*}

\paragraph{Prior mismatch and value drift} The last two blocks of \cref{S-tab:r049} report the fresh-secret gains when the attacker's prior table comes from 2{,}000 firms of another population model, with the evaluated firms unchanged, and when prices and costs drift between queries. PM1 uses Student-$t$ price noise with 30 instead of 3 degrees of freedom, multiplies the spreads of the firm premium, the city residual and the firm efficiency by 1.5, and has no extreme firms; PM2 multiplies every class price median by $e^{N(0,0.25^2)}$ and shifts every sector margin median by $N(0,0.3^2)$ on the logit scale, drawn once; PM3 combines both. Under the three prior models the single-query margin advantages are +0.6 to +1.0\pp under \Donep, +2.0 to +4.3\pp under \Dtwo (intervals above zero under PM2 and PM3) and +2.1 to +2.8\pp under \Dthreeo at $\varepsilon=1$. \Done's reuse unit-cost gain is +8.9, +3.9 and +7.8 [0.6, 15.6]\pp under PM1--PM3, against fresh gains above +41\pp, and the \Donep and \Dtwo reuse gains are zero; under PM3 the joint attacker gains +32.8 [20.6, 44.4]\pp with fresh maps and +1.7 [$-$2.2, 6.1]\pp with one reused map. Under drift the frozen attacker treats every view as exact, and a sequence that no prior sample (or key) can explain reverts to the prior: at $\delta=5\%$ this happens in 20 to 22 of 180 \Dthreeo sequences, 11 \Donep sequences and 2 \Done sequences.

\paragraph{Reference likelihood} Replacing the Gaussian likelihood with a kernel density over 20{,}000 simulated maps (\cref{tab:d2valid}) raises the default-law margin gain from +34.4 to +43.3 [33.3, 53.3]\pp, from 26.1\% to 69.4\%, and makes the unit-cost gain significant (+7.8 [1.1, 14.4]\pp); under Dirichlet(0.1) gaps it lowers the gain from +12.2 to +8.3 [1.7, 15.6]\pp, still above zero. The gain of \cref{fig:linked} is therefore conservative relative to this reference at $n=16$, although the two curves cross at intermediate $n$. Across the bandwidths and map counts we tried, its hit rates moved by at most 1.8\pp.

\paragraph{Joint attacker} With four times as many Monte Carlo draws or a smoothing of 0.1 instead of 0.03, the fresh-minus-reuse margin difference is +30.6 [20.0, 41.7] and +34.4 [23.3, 45.6]\pp, with reuse gains between $-2.2$ and +0.6\pp; a smoothing of 0.01, which keeps more of the within-interval relations, gives +31.7 and +33.9\pp with reuse gains of at most +5.6\pp (\cref{S-tab:numval}).

\begin{table*}[!t]
\centering
\footnotesize
\setlength{\tabcolsep}{5pt}
\caption{Numerical Validation of the \Dtwo Reuse-Side Attackers}
\label{S-tab:numval}
\begin{tabular}{@{}llrrr@{}}
\multicolumn{5}{@{}l}{\emph{(a) Joint shared-map attacker: smoothing and Monte Carlo draws}} \\
\toprule
Smoothing (log units) & Draws & Fresh margin gain & Reuse margin gain & Fresh $-$ reuse \\
\midrule
0.03 & 400 (main) & +30.0 [+18.9, +40.6] & +0.6 [-4.4, +6.1] & +29.4 [+18.3, +40.6] \\
0.03 & 1,600 & +31.1 [+20.0, +42.2] & +0.6 [-4.4, +5.6] & +30.6 [+20.0, +41.7] \\
0.1 & 400 & +32.2 [+21.7, +42.8] & -2.2 [-8.9, +4.4] & +34.4 [+23.3, +45.6] \\
0.01 & 400 & +37.2 [+27.2, +47.2] & +5.6 [-1.1, +12.2] & +31.7 [+21.7, +41.7] \\
0.01 & 1,600 & +37.8 [+27.8, +47.8] & +3.9 [-1.7, +9.4] & +33.9 [+23.9, +43.9] \\
\bottomrule
\end{tabular}

\vspace{6pt}
\begin{tabular}{@{}lrrr@{}}
\multicolumn{4}{@{}l}{\emph{(b) A-exact on 60 reused-map sequences (first 20 test firms $\times$ 3 seeds): computational budget}} \\
\toprule
Particles / initial maps / moves & $n{=}1$ & $n{=}16$ & Reuse margin gain \\
\midrule
20{,}000 / 400{,}000 / 40 (main settings) & 31.7 & 35.0 & +3.3 [-13.3, +18.3] \\
80{,}000 / 1{,}600{,}000 / 160 & 31.7 & 33.3 & +1.7 [-15.0, +18.3] \\
\bottomrule
\end{tabular}
\par\vspace{2pt}\parbox{\textwidth}{\footnotesize\raggedright Margin hits ($\pm5$\,pp), original firms, D2 default law; gains from $n{=}1$ to 16 in pp with firm-clustered 95\% bootstrap CIs; (a) 60 test firms $\times$ 3 seeds, margin hit rates in \%. In (b), the paired budget difference in the reuse gain is $-1.7$ [$-$5.0, 0.0]\,pp and the mean absolute difference of the margin estimates is 0.003 ($n{=}1$) and 0.004 ($n{=}16$). Rejection sampling, the exact reference used for small one-digit instances (\cref{app:attack}), accepts almost no draws at six significant digits.}
\end{table*}

\paragraph{A-exact} The main estimates use sampler seed 4646. With seed 4647 the reuse gains are +3.3\pp on the original firms and $-1.1$\pp on the fresh draw (seed 4646: +5.6 and +0.0\pp); the two seeds' margin hit rates differ by at most 2.2\pp and their unit-cost hit rates by at most 1.1\pp at $n=1$ and 16 under the default law, and by at most 3.3\pp over all evaluated $n$, map laws and single-query subsets. Set against the joint attacker's fresh-map gain, A-exact's reuse gain gives differences of +24.4 [9.4, 38.3]\pp (fresh draw +22.8 [7.8, 37.2]\pp), and +26.7 and +23.9\pp with the second seed. Under the other map laws of \cref{S-tab:d2law} its reuse margin gain is +3.9 [$-$2.2, 10.6]\pp (Dirichlet(0.1) gaps), +8.3 [$-$2.8, 19.4]\pp (16 intervals) and +2.2 [$-$5.0, 9.4]\pp (256 intervals), with every interval including zero in both seeds (second seed +2.2, +6.7 and +2.2\pp). A-exact is only approximately converged: the 10\% quantile over sequences of its smallest per-update effective sample size is about 12, with 400{,}000 maps at the first query, and rejection sampling, the exact reference, accepts almost no draws at six significant digits. On an Apple M4 with eight worker processes it takes about one to two minutes per 16-query sequence under the default law and about four minutes with 256 intervals.

\paragraph{Map law} \Cref{S-tab:d2law} varies the law, with the attacker told each law. The margin gain from $n=1$ to 16 stays significant under every law we tried (+12.2 to +51.1\pp) and grows as the map becomes more regular; under the most regular law the single-query margin advantage is already +28.0 [23.6, 32.5]\pp. Widening the target range to $[1,10^{12}]$ leaves the gain at +35.6 [25.0, 46.7]\pp (default +34.4). The least regular law (Dirichlet(0.1) gaps) yields the smallest gain, +12.2 [1.7, 22.2]\pp, with a single-query margin advantage of $-3.0$ [$-$7.3, 1.3]\pp.

\begin{table*}[!t]
\centering
\small
\setlength{\tabcolsep}{4pt}
\caption{\Dtwo under Alternative Map Laws}
\label{S-tab:d2law}
\begin{tabular}{@{}lrrrrr@{}}
\toprule
 & \multicolumn{2}{c}{Single query, $\Delta$ vs \Azero} & \multicolumn{3}{c}{Linked, margin hit (\%)} \\
\cmidrule(lr){2-3}\cmidrule(lr){4-6}
Map law & margin & cost & $n{=}1$ & $n{=}16$ & gain \\
\midrule
default: 64 intervals, uniform cuts, $[1,10^6]$ & +1.0 [-2.7, +4.5] & -0.2 [-3.0, +2.5] & 25.0 & 59.4 & +34.4 [+23.9, +45.0] \\
16 intervals & +0.7 [-3.1, +4.4] & +0.7 [-2.2, +3.5] & 22.2 & 56.7 & +34.4 [+22.2, +46.1] \\
256 intervals & +4.6 [+0.9, +8.3] & +1.2 [-1.8, +4.2] & 22.2 & 71.7 & +49.4 [+38.3, +60.0] \\
irregular gaps, Dirichlet(0.1) & -3.0 [-7.3, +1.3] & -1.1 [-4.3, +2.1] & 17.2 & 29.4 & +12.2 [+1.7, +22.2] \\
regular gaps, Dirichlet(10) & +28.0 [+23.6, +32.5] & +3.5 [+0.1, +6.9] & 46.1 & 97.2 & +51.1 [+42.2, +60.0] \\
targets $[1,10^{12}]$ & +0.8 [-2.9, +4.4] & -0.3 [-3.1, +2.4] & 23.9 & 59.4 & +35.6 [+25.0, +46.7] \\
\bottomrule
\end{tabular}
\par\vspace{2pt}\parbox{\textwidth}{\footnotesize\raggedright The attacker is told each law (it simulates its Gaussian moments for that law; solver otherwise frozen). Single query: advantage of \Aplus over \Azero on the test split (2{,}100 rows). Linked: fresh maps, product~0, 60 test firms $\times$ 3 seeds; margin hit rate at $n{=}1$ and $16$ and the gain. 95\% CIs (pp). The default law reproduces \cref{tab:single} and \cref{fig:linked}.}
\end{table*}

\section{Language-Model Study}
\label[appendixsec]{S-sec:llm}
\paragraph{Attackers} We tested three attackers with Llama-3.1-8B-Instruct~\cite{grattafiori2024llama3} and with Qwen2.5-32B-Instruct under 4-bit AWQ quantization~\cite{qwen2025qwen25}, served with vLLM~\cite{kwon2023vllm} and greedy decoding. A2 reads the prose with numbers masked and guesses margins and costs; A3 reads the whole disguised prompt plus a description of the mechanism; A4 reads the masked prose and returns the sector, the product classes, the city's price level on a scale where the US average is 100, the percentage price effect of the positioning and a margin shift. The city index $I\in[50,200]$ and positioning effect $q\in[-60,150]$\% (values outside are clipped) give a price multiplier $\min(\max(\tfrac{I}{100}(1+\tfrac{q}{100}),0.25),4)$, whose logarithm shifts the centre of each product's log-price prior; the margin shift $s\in[-40,40]$\pp is added to the class median margin (clipped to $[0.01,0.97]$), and the frozen \Aplus solver then runs with unchanged prior widths. With the true sector and classes and no shift, A4 reduces to \Aplus exactly. Prompts were tuned on the development split only and request structured JSON; an A2 or A3 answer that cannot be parsed after three attempts counts as a miss, and a failed A4 extraction falls back to \Aplus without the oracle and is flagged.

\paragraph{Prespecified test} Before running the test split we recorded in a version-controlled plan two decision endpoints, C1a (unit-cost hits under \Done) and C1b (unit-cost or margin hits under \Dthree at $\varepsilon=0.5$), and the success criterion: a positive A4$-$\Aplus difference with a confidence interval above zero on the same applicable endpoint for at least two models. The test split was run once for both models; \cref{S-tab:llm} reports the recorded answers, replayed from cache to add \Dthreeo.

\begin{table*}[!t]
\centering
\footnotesize
\setlength{\tabcolsep}{3.5pt}
\caption{Language-Model Priors: Differences to \Aplus}
\label{S-tab:llm}
\begin{tabular}{llrrrrrr}
\toprule
 & & \multicolumn{2}{c}{Llama-3.1-8B} & \multicolumn{2}{c}{Qwen2.5-32B} & \multicolumn{2}{c}{Oracle cues (no LLM)} \\
\cmidrule(lr){3-4}\cmidrule(lr){5-6}\cmidrule(lr){7-8}
Mechanism & Metric & A4 & A4-TL & A4 & A4-TL & city & city + premium \\
\midrule
D1 & cost & +1.9 [-3.4, +7.0] & +2.3 [-3.0, +7.4] & -2.3 [-7.9, +3.2] & -2.4 [-7.9, +3.1] & -1.9 [-6.1, +2.0] & +4.0 [-1.5, +9.3] \\
D2 (ours) & margin & -0.1 [-1.6, +1.3] & 0.0 [-1.5, +1.4] & +1.0 [-1.0, +3.1] & +1.0 [-1.0, +3.1] & +0.3 [0.0, +0.7] & +0.5 [+0.1, +1.0] \\
D2 (ours) & cost & -0.2 [-2.9, +2.5] & +0.2 [-2.4, +2.9] & -2.4 [-5.2, +0.3] & -2.4 [-5.2, +0.3] & +0.2 [-2.1, +2.7] & +1.9 [-1.4, +5.3] \\
D3, $\varepsilon$=0.5 & cost & +2.4 [+0.7, +4.3] & +2.5 [+0.8, +4.3] & +2.0 [-0.1, +4.0] & +1.9 [-0.2, +3.9] & +2.3 [+0.8, +4.0] & +1.0 [-1.3, +3.3] \\
D3, $\varepsilon$=0.5 & margin & -0.9 [-2.5, +0.7] & -0.9 [-2.5, +0.6] & -0.6 [-2.7, +1.5] & -0.6 [-2.7, +1.5] & 0.0 [-0.9, +1.1] & -1.0 [-2.7, +0.4] \\
D3o, $\varepsilon$=1 & cost & +0.7 [-1.5, +2.9] & +0.9 [-1.3, +3.0] & -1.2 [-3.8, +1.2] & -1.2 [-3.8, +1.2] & -- & -- \\
D3o, $\varepsilon$=1 & margin & -0.2 [-2.2, +1.8] & -0.2 [-2.2, +1.8] & +1.2 [-1.6, +4.0] & +1.2 [-1.6, +4.0] & -- & -- \\
\bottomrule
\end{tabular}
\par\vspace{2pt}\parbox{\textwidth}{\footnotesize\raggedright Paired differences in hit rate (pp; cost $\pm$10\%, margin $\pm$5\,pp) with firm-clustered 95\% bootstrap CIs (10{,}000 resamples; 2{,}000 for oracle cues); test split, 2{,}100 rows per cell. A4: hybrid that reads the masked prose and shifts the prior (recorded test-run answers, replayed from cache). A4-TL: same cue shifts with the true sector and class. Oracle cues: the generator's true city level (and brand premium) fed to the same solver, no LLM.}
\end{table*}

\paragraph{Results} On the development prompts the models' city price levels correlate with the regional price parities (Spearman 0.63 for Llama and 0.53 for Qwen), and they assign lower prices to discount firms and higher prices to premium firms. In our generator, however, the city explains little of a product's price relative to the spread within a product class: feeding the true city level into the same solver gives +2.3 [0.8, 4.0]\pp on \Dthree $\varepsilon=0.5$ unit cost, and adding the true brand premium gives between $-1.0$ and +4.0\pp across endpoints, the largest with interval [$-$1.5, 9.3]\pp (\Done unit cost). A4 gained +1.9 [$-$3.4, 7.0]\pp (Llama) and $-$2.3 [$-$7.9, 3.2]\pp (Qwen) on C1a, and +2.4 [0.7, 4.3]\pp (Llama) and +2.0 [$-$0.1, 4.0]\pp (Qwen) on the C1b unit-cost endpoint; margin differences on C1b were negative, and no difference is significant under \Dthreeo. Only one model had a gain with an interval above zero, so the prespecified two-model criterion was not met. Replacing the models' sector and class answers by the truth while keeping their cue shifts (A4-TL) changes any difference by at most 0.43\pp; the largest observed gain is +2.5 [0.8, 4.3]\pp (A4-TL, Llama, \Dthree $\varepsilon=0.5$ unit cost).

\paragraph{Direct-generation baselines} A2 and A3 answer from the prompt without the structured solver. Under the disguised mechanisms A2 hits fewer margins than the prior-only attacker (17.7\% and 20.4\% versus 28.1\%). A3 computes margins from plaintext (81.2\% and 85.6\%) and partly uses the shared-scaling invariant (39.0\% and 86.9\% margin hits under \Done, where \Aplus reaches 100\%), but stays below \Azero under \Donep and \Dtwo; it returned no usable answer in at most 125 of 2{,}100 rows per cell.

\bibliographystyle{IEEEtran}
\bibliography{references}

\end{document}